\documentclass[%
reprint,
aps,
pra,
twocolumn,
amsmath,amssymb,bibnotes
]{revtex4-1}
\usepackage{graphicx}
\usepackage{dcolumn}
\usepackage{bm}
\usepackage{dcolumn}
\usepackage{dsfont}
\usepackage{bbm}
\usepackage{hyperref}
\usepackage[mathlines]{lineno}
\usepackage{natbib}
\usepackage{cases}

\begin{document}
\title{Topological Edge States in Nanoparticle Chains: Isolating Radiative Heat Flux}
\author{M.~Nikbakht}
\email{mnik@znu.ac.ir}
\author{F.~Bahmani}
\affiliation{Department of Physics, University of Zanjan, Zanjan 45371-38791,Iran.}
\date{\today}
\begin{abstract}
Recent advancements in the field of topological band theory have significantly contributed to our understanding of intriguing topological phenomena observed in various classical and quantum systems, encompassing both wave and dissipative systems. In this study, we employ the notion of band theory to establish a profound connection between the spatio-temporal evolution of temperatures and the underlying topological properties of radiative systems. These systems involve the exchange of energy through radiation among a collection of particles.  By utilizing the eigenstates of the response matrix of the system, we establish a robust framework for the examination of topological properties in radiative systems, considering both symmetric and asymmetric response matrices. Our formalism is specifically applied to investigate the topological phase transition in a one-dimensional chain composed of an even number of spherical nanoparticles. We provide compelling evidence for the existence and robustness of topological edge states in systems characterized by an asymmetric response matrix. Moreover, we demonstrate that by manipulating the arrangement and volume of particles, it is possible to control the system's structure and achieve desired topological features. Interestingly, we showed that the radiative heat transfer can be controlled and prevented by topological insulation. Additionally, we conduct an analysis of the temperature dynamics and the associated relaxation process in the proposed system. Our research findings demonstrate that the interplay between bulk states and localized states is pivotal in the emergence of distinct eigenstates and provides significant insights into the spatio-temporal dynamics of temperature and the process of thermalization within a system. This interplay holds the potential to be leveraged for the development of structures, facilitating efficient heat transfer even in the presence of perturbation. Consequently, it enables precise experimental measurements of heat transfer and serves as a platform for the exploration of thermal topology, offering new avenues for scientific inquiry in this field.

\end{abstract}
\maketitle

It is widely recognized that when two objects are in close proximity (compared to the thermal wavelength $\lambda_T=2\pi\hbar c/k_BT$), the radiative heat exchange between them is significantly larger than what is predicted by Stefan-Boltzmann's law~\cite{PhysRevB.4.3303}. As a result, the thermal transport and thermalization processes in a system of nanoparticles (NPs) are intricately linked to factors such as their geometrical arrangement, shape, composition, and inter-particle separation distances~\cite{RevModPhys93025009,Nikbakht2015,PhysRevB102024203,WEN20191,CZAPLA20194,NIKBAKHT2018164,PhysRevB96155402,PhysRevB106195417,Choubdar}.

Theoretical advancements and experimental measurements have spurred significant efforts in thermal management and regulation within particle ensembles~\cite{PhysRevB105205422,PhysRevB95125411,10108813616633abe52b,PhysRevLett107114301,manybody}, as well as structures with planar geometries~\cite{PhysRevB.95.205404,song2015near,xu2019near}. Over the past few years, numerous studies have focused on developing mechanisms to control the magnitude and direction of radiative heat exchange. The recent progress in passive and active control of radiative heat transfer has generated considerable interest in phenomena such as thermal rectifiers~\cite{PhysRevB104045410,LI2022100632,HUANG2023123942,PhysRevB106075408}, magneto-resistance~\cite{PhysRevLett118173902, PhysRevB101085411}, persistent heat fluxes~\cite{PhysRevB97205414,PhysRevLett117134303}, thermal barriers~\cite{PhysRevB102035433}, heat shuttling\cite{PhysRevLett.121.023903}, orientation-based regulators~\cite{doi10106350018329, HU2022122666}, material-based regulators~\cite{graphene,CHEN2023108540,doi10106350087089,PhysRevApplied19024019}, heat pumping\cite{PhysRevB.101.165435,PhysRevApplied.4.011001}, and thermal switching~\cite{thompson2019nanoscale,gu2015thermal,YU2022123339,Yu22}.

Recent studies have provided valuable insights into the influence of topology on dipolar interactions, topological invariance, and the resilience of localized modes in systems composed of particles with dipolar interaction. Downing and Weick conducted research on the topological properties of collective plasmonic excitations in bipartite chains of spherical metallic nanoparticles, shedding light on the interplay between topology and dipolar interactions~\cite{PhysRevB.95.125426, Downing2018}. Ling et al. investigated the existence of topological edge plasmon modes in diatomic chains consisting of identical plasmonic nanoparticles, further elucidating their emission rate~\cite{Ling:15}. Pocock et al. performed a comprehensive analysis of wave systems, emphasizing the longitudinal and transverse modes, as well as the significance of disorder on the topological properties~\cite{acsphotonics.8b00117}. Wang and Zhao focused on the topological phases exhibited by one-dimensional dimerized doped silicon nanoparticle chains, providing crucial insights into the interplay between topology and material properties~\cite{1.5131185}. Nikbakht and Mahdieh examined the localization of dipolar modes in fractal structures~\cite{jp1088542}. 
In addition to the conventional studies mentioned earlier, recent research endeavors are also investigating the influence of topology on radiative heat exchange in the static regime. For instance, Nikbakht investigated the localization of thermal modes in fractal structures~\cite{PhysRevB96125436}. Additionally, Ott and Biehs investigated the thermal topology in plasmonic InSb nanoparticles~\cite{PhysRevB102115417,OTT2022122796}. Luo et al. proposed a topological insulator analog in radiative systems~\cite{LUO2023100984}. Ott et al. studied the local density of states in plasmonic nanostructures.~\cite{PhysRevB104165407}. Herz and Biehs proposed thermal radiation and near-field thermal imaging of a plasmonic Su-Schrieffer-Heeger chain~\cite{doi10106350123515}. However, it is important to note that these non-trivial topological behaviors, arising from the localization of dipolar modes in the systems, may not have a significant impact on transient or steady-state heat transfer in the system.

On the basis of recent theoretical studies, the temporal evolution of temperatures in collection of particles can be simplified by linearizing the energy balance equation~\cite{PhysRevLett126193601,PhysRevB104024301,DONG2022123318}. In these approaches, the dynamic of temperatures  becomes a simple eigenvalue problem.  From the dynamical perspective, in addition to the features we mentioned earlier, the radiative heat transfer between NPs will be affected by spatio-temporal coupling of the temperatures. Thus, the temporal evolution of temperatures is not solely determined by the radiative heat exchanges, but also by the history dependent dynamics. Several theoretical and experimental  works based on topological band theory has revealed that most of the unusual behaviors in the dynamics of wave systems can be governed by their topological properties~\cite{PhysRevLett481559,PhysRevLett612015,Nakajima2016,PhysRevLett49405}. Unlike quantum mechanical or classical wave systems, thermal evolution in radiative systems is of dissipative nature. Recent studies have suggested that asymmetric analogues of the Su-Schrieffer-Heeger (SSH) model are relevant in describing one-dimensional topological behavior in dissipative systems~\cite{Cao2021,Yoshida2021sc,Diffusion,Xu2022}. Ultimately, it will be of interest to propose a system showing topological features like insulation in radiative systems.

In this study, we apply the concept of band topology, which is traditionally used in classical and quantum wave systems, to investigate heat transfer in radiative systems. We demonstrate how the utilization of the response matrix allows for an understanding of the topological behavior exhibited by these systems. Our calculations are carried out within the theoretical framework of fluctuational electrodynamics, specifically in the dipolar approximation regime. By employing a linearization approach, we determine the eigenvalues and eigenstates of the response matrix, which govern the temporal evolution of temperatures in the underlying radiative system. The topological phases observed are then interpreted in terms of these eigenstates. We utilize this methodology to illustrate a topological phase transition and the emergence of localized states with topological robustness in a chain of spherical nanoparticles. Subsequently, we investigate the thermalization process in the presence of localized states and unveil the existence of a simple topological insulator in radiative systems.

\section{Theoretical model}  
Consider a bipartite chain of spherical NPs along the $x$ axis with lattice constant $d$, as depicted in Fig.~(\ref{Figure.1}a). The intra-cell separation distance  between NPs A and B is $\beta d$, and volume of NPs are $V_i(\beta)$. Here, the number of particles is taken to be even and $\beta\in [0.45,0.55]$ is used as  topological tuning parameter. Particles are initially at temperatures $T_i$ and immersed in a thermal bath at constant temperature $T_b=300$~K. Following previous studies, the differential equation governing the evolution of  temperatures is~\cite{PhysRevB.88.104307,Nikbakht_2015}
\begin{equation}
\gamma_i\frac{d T_i}{d t}=\mathcal P_i,~~(i=1,2,\cdots,N),
\label{eq1}
\end{equation}
where  $\mathcal{P}_i$ refers to the radiative  power dissipated in particle $i$, respectively.  Moreover, $\gamma_i=c_iV_i$  is the heat capacity of the particle, with volumetric specific heat capacity $c_i$. Within the linear approximation we get a Schrodinger $\emph{like}$ equation, which takes an elegant form when written in compact notation
\begin{equation}
\frac{d}{dt}\Delta{ {\bf T}}(t)=-{\hat H}\Delta{ {\bf T}}(t),
\label{eq2}
\end{equation}
where $\Delta{\mathbf{T}}=(\Delta T_1,\Delta T_2,\cdots,\Delta T_N)^\intercal$ is the column vector representing the temperature state, with elements $\Delta T_i=T_i-T_b$ that describe the temperature deviation of the NPs with respect to the environment. Moreover, ${\hat H}=-\hat\Gamma^{-1}{\hat F}$ is $N\times N$ $\emph{static}$ response matrix with elements $H_{ij}=-F_{ij}/\gamma_i$, and $\hat \Gamma$ is a diagonal matrix with  heat capacities along the diagonal, i.e., $\hat\Gamma=\text{diag}\{\gamma_1,\gamma_2,\cdots,\gamma_N\}$. Following previous works~\cite{PhysRevLett126193601,PhysRevB104024301,DONG2022123318}, the elements of the conductance matrix are given by
\begin{equation}
F_{ij}=\int_0^\infty \frac{d\omega}{2\pi} \tau_{ij}(\omega)\frac{\partial \Theta(\omega,T)}{\partial T}{\Big |}_{T_b}.
\label{eq3}
\end{equation}
In this equation, $\Theta(\omega,T)=\hbar\omega/[\exp (\hbar\omega/k_BT)-1]$ is the  mean energy of a Planck harmonic oscillator in thermal equilibrium at frequency $\omega$ and temperature $T$. Moreover, $\tau_{ij}=2{\tt Im}{\mathrm {Tr}}[{\hat{\mathbb A}}_{ij}{\tt Im}({\hat {\bm \chi}}_{j}){\hat{\mathbb C}}_{ij}^\dag]$ represents the radiative energy transmission coefficient between $i$-th and $j$-th NPs, which includes many-body effects (for more details see Refs.~\cite{manybody,PhysRevLett107114301,PhysRevB95125411}). In this expression $3N\times 3N$ block matrices are defined as ${\mathbb A}=[\mathbbm{1}-\widetilde{\alpha}{\mathbb W}]^{-1}$, $\hat{\bm \chi}=\widetilde{\alpha}+\widetilde{\alpha}{\hat{\mathbb G}}_0^\dag\widetilde{\alpha}^\dag$, and $\hat{\mathbb C}=[\mathbb{W}+{\hat{\mathbb G}}_0]\mathbb A$. Here, $\mathbbm{1}$ stands for the identity matrix, $\hat{\mathbb W}$ is the dipole-dipole interaction matrix, ${\hat{\mathbb G}}_0=i(k^3/6\pi)\mathbbm{1}$, and $\widetilde{\alpha}=\text{diag}\{\hat{\bm \alpha}_1,\hat{\bm\alpha}_2,\cdots,\hat{\bm\alpha}_N\}$ is the polarizability matrix. Furthermore, the $3\times 3$ block matrices in ${\mathbb W}$ represent the dipolar interaction between particles $i$ and $j$. This interaction is described by the free space Green's tensor $\hat{\mathbb G}_{i\neq j}=\frac{k^3}{4\pi}\Big[f(kr_{ij})\mathbbm{1}+g(kr_{ij})\frac{{\bf r}_{ij}\otimes{\bf r}_{ij}}{r_{ij}^2}\Big]$. In this equation, $f(x)=[x^{-1}+ix^{-2}-x^{-3}]\exp(ix)$ and $g(x)=[-x^{-1}-3ix^{-2}+3x^{-3}]\exp(ix)$, where $k=\omega/c$ and $r_{ij}$ represents the distance between the $i$-th and $j$-th nanoparticles, which are located at positions ${\bf r}_{i}$ and ${\bf r}_{j}$, respectively.

In the case of identical particles, the heat capacity matrix simplifies to $\hat{\Gamma} = \gamma \mathbbm{1}$, where $\gamma$ represents the heat capacities. Furthermore, since $\hat{F}$ is symmetric, the response matrix becomes symmetric. However, in a system of NPs with different volumes, the heat capacity matrix takes the form $\hat\Gamma=c~\text{diag}\{V_1,V_2,\cdots,V_N\}$. As a result, the response matrix becomes asymmetric. Nonetheless, the response matrix remains diagonalizable, allowing us to express it in biorthonormal form as $\hat H=\sum_{\mu=1}^{N}\lambda_\mu\boldsymbol{\psi}_\mu \boldsymbol{\phi}_\mu^\intercal$. Here, $\boldsymbol{\psi}_\mu$ and $\boldsymbol{\phi}_\mu$ represent the right and left eigenvectors of $\hat H$, respectively, corresponding to the eigenvalue $\lambda_\mu$. They satisfy the equations ${\hat H}\boldsymbol{\psi}_\mu=\lambda_\mu\boldsymbol{\psi}_\mu$ and $\boldsymbol{\phi}_\mu^\intercal{\hat H}=\lambda_\mu\boldsymbol{\phi}_\mu^\intercal$. Throughout the paper, the vector $\boldsymbol{\psi}_\mu$ along with its corresponding eigenvalue is referred to as the $\mu$th mode. Additionally, for convenience, we use indices $\{i, j, m, n\}$ to label the nanoparticles (i.e., lattice sites), and Greek indices $\mu$ and $\nu$ to denote the mode numbers.

It is important to emphasize that in general, $\boldsymbol{\psi}_\mu$ are not mutually orthogonal, and $\boldsymbol{\psi}_\mu \neq \boldsymbol{\phi}_\mu$. However, due to the diagonalizability of the response matrix, its left and right eigenvectors satisfy biorthogonality, i.e., $\boldsymbol{\phi}_\mu^\intercal \cdot\boldsymbol{\psi}_\mu = \delta_{\mu\nu}$. Additionally, it can be shown that these eigenvectors are related by $\boldsymbol{\phi}_\mu = \hat \Gamma \boldsymbol{\psi}_\mu / (\boldsymbol{\psi}_\mu^\intercal \hat \Gamma \boldsymbol{\psi}_\mu)$.

To examine the temporal changes in temperatures, it is valuable to examine the continuum basis denoted by ${\psi_\mu(x)}$, where $\psi_\mu(x)$ represents the eigenstate of the Hamiltonian operator $\hat H$ in the spatial representation. Likewise, the temperature distribution $\Delta T(x,t)$ can be referred to as the state of temperature. Through meticulous calculations (outlined in Section (\ref{sec2}) of the appendix), the solution to Equation (\ref{eq2}) is derived as follows:

\begin{equation}
\Delta{T}(x,t) = \sum_{\mu=1}^{N} C_\mu(0) e^{-\lambda_\mu t} \psi_\mu(x),
\label{eq4}
\end{equation}

where $C_\mu(0) = \boldsymbol{\phi}_\mu^\intercal \cdot \Delta{\mathbf{T}}(0)$ represents the initial weight of the temperature state in the $\mu$th mode. This equation provides the temperature state as a function of spatial and temporal coordinates. By knowing the initial temperature state, we can calculate the initial weight distribution as $P_\mu = |\boldsymbol{\phi}_\mu^\intercal \cdot \Delta{\mathbf{T}}(0)|^2 / [\Delta{\mathbf{T}}(0)^\intercal \cdot \Delta{\mathbf{T}}(0)]$. In the supplementary file, we discussed how the weight of the initial temperature state vector plays a significant role in the thermalization process of the system.

\section{Localized modes in radiative systems} 

\begin{figure}
\includegraphics[scale=1]{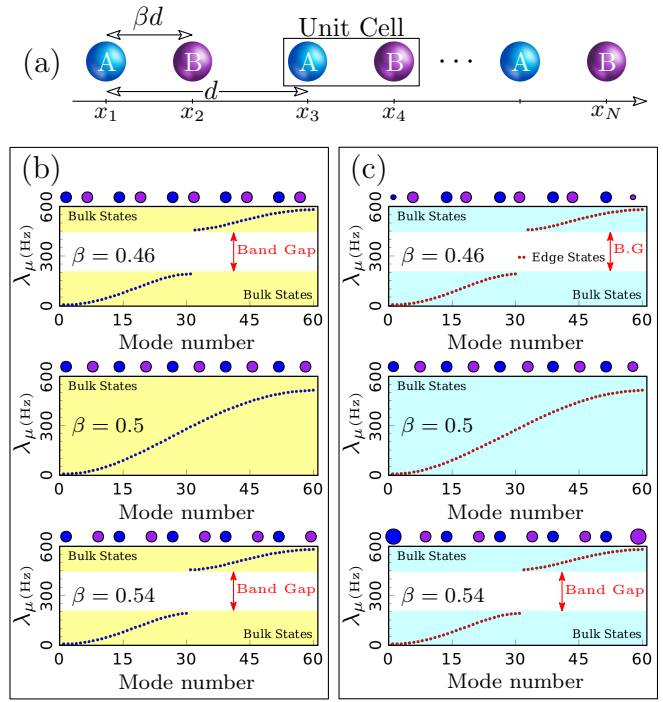}
\caption{(a) A schematic representation of a bipartite chain of spherical nanoparticles (NPs) aligned along the $x$ axis with a lattice constant of $d$. The NPs are divided into $A$ and $B$ sub-lattices, with an intra-cell separation distance of $\beta d$. The chain is immersed in a thermal bath at a temperature of $T_b=300$ K.
(b) Eigenvalue spectrum of the SSH chain with a symmetric response matrix, showcasing typical values of $\beta\in\{0.46,0.5,0.54\}$. The lattice constant is $d=500$ nm, the number of particles is $N=60$, and the volumes of the NPs are constant at $V_i=V_0=\frac{4\pi}{3}(50)^3$ nm$^3$.
(c) Eigenvalue spectrum of the SSH chain with an asymmetric response matrix. The parameters are the same as in panel (b), except for the volume of particles 1 and 60, which depend on $\beta$ according to $V_1(\beta)=V_{60}(\beta)=\frac{4\pi}{3}[1250000\beta-561000]$ nm$^3$.}
\label{Figure.1}
\end{figure}

The topological properties of the SSH chain shown in Fig.(\ref{Figure.1}a) are determined by the eigenvalue spectrum of its response matrix. In the case of reciprocal nanoparticles, the eigenvalues of the response matrix are real and depend solely on the tuning parameter $\beta$. In this study, we focus specifically on the thermal topology of systems consisting of reciprocal nanoparticles. As a concrete example, we consider Silicon-Carbide (SiC) as a typical material, characterized by a volumetric specific heat capacity of $c=2.4075\times 10^6\text{J}\text{K}^{-1}\text{m}^{-3}$. The polarizability of the nanoparticles is given by $\alpha_i=3V_i(\epsilon-1)/(\epsilon+2)$, where $\epsilon$ represents the scalar permittivity defined as $\epsilon=\omega_\infty(\omega_L^2-\omega^2-i\gamma\omega)/(\omega_T^2-\omega^2-i\gamma\omega)$. Here, we adopt specific values for the SiC material, namely $\epsilon_\infty=6.7$, $\omega_T=1.495\times10^{14}\text{rad/s}$, $\omega_L=1.827\times10^{14}\text{rad/s}$, and $\gamma=0.9\times10^{12}~\text{rad/s}$.

A simple example of the radiative analogue of the SSH model is a chain of nanoparticles with identical volumes. In this case, the parameter $\beta$ only affects the intra-cell separation distance, resulting in a symmetric response matrix. In Fig.~(\ref{Figure.1}b), we present the eigenvalue spectrum of the system for $N = 60$ nanoparticles with a lattice constant of $d=500$ nm. The nanoparticles have a constant volume $V_i=V_0=\frac{4}{3}\pi(50)^3$ nm$^3$, and the results are shown for typical values of $\beta$ in the range $\beta\in\{0.46,0.5,0.54\}$.
\begin{figure}
\includegraphics[scale=1]{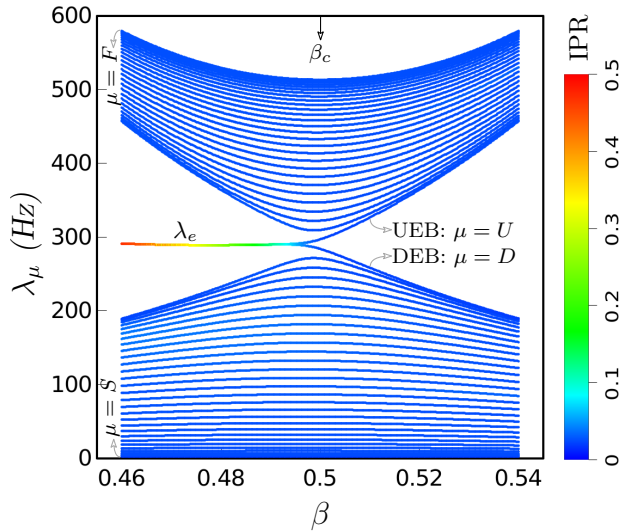}
\caption{The eigenvalue spectrum of the response matrix $H$ is shown as a function of $\beta$ for an SSH chain consisting of $N = 60$ SiC nanoparticles with a lattice constant of $d=500$ nm. The volume of the first and last nanoparticles depends on $\beta$ according to $V_1(\beta)=V_{60}(\beta)=\frac{4\pi}{3}[1250000\beta-561000]$ nm$^3$, while all other nanoparticles have constant volumes $V_{i\neq 1,60}=\frac{4\pi}{3} 50^3$ nm$^3$. The modes are sequentially labeled using Greek indices, and the key bands include $S$ (slowest band), $F$ (fastest band), $U$ (higher localized band), and $D$ (lower localized band). The color bar on the right represents the Inverse Participation Ratio (IPR) of the eigenstates.}
\label{Figure.2}
\end{figure}
The eigenvalue spectrum reveals a prominent gap at $\beta=0.46$, indicating a substantial separation between relaxation rate levels. However, at the critical value $\beta=\beta_c=0.5$, the band gap in the spectrum closes, suggesting a transition in the system's states. Subsequently, for values of $\beta>\beta_c$, as exemplified by $\beta=0.54$, the gap reopens, indicating a similar relaxation configuration. To comprehensively assess the topological properties of the system, we calculate the Zak phase and winding number (refer to section~(\ref{Zakandwinding}) in the Appendix). Our calculations demonstrate that the structure undergoes a topological phase transition across the critical value of $\beta=0.5$. Importantly, this transition does not coincide with the emergence of localized states within the system.

\begin{figure*}[t]
\includegraphics[]{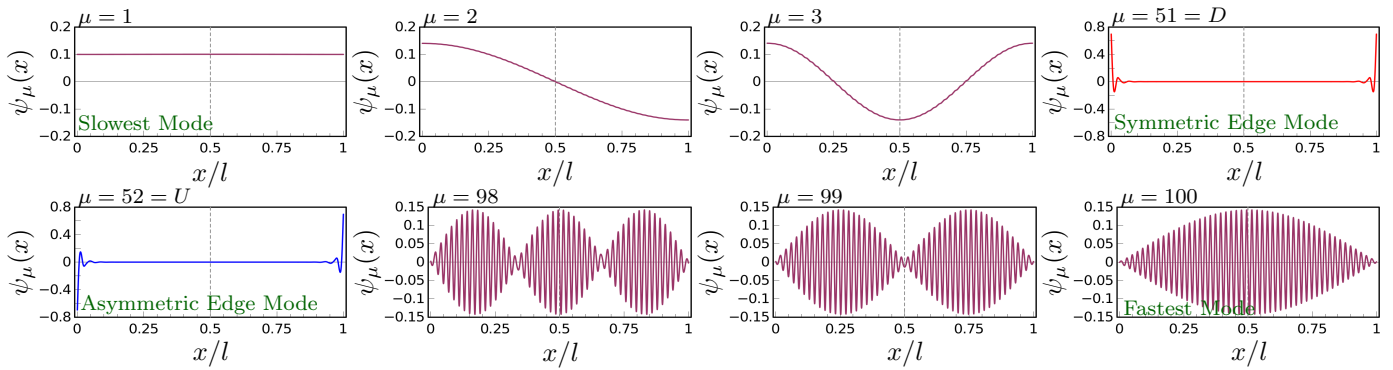}
\caption{The eigenmode profiles are shown for selected modes of the response matrix in a chain of $N = 100$ nanoparticles with an inter-particle distance of $d = 500$ nm. These profiles specifically correspond to the case of $\beta=0.46$, representing the system in its topologically nontrivial phase. In this setup, the volumes of the nanoparticles, except for the first and last ones, are kept constant at $V_{i\neq 1,100}=\frac{4}{3}\pi (50)^3 \text{nm}^3$. However, the first and last particles, denoted as $V_1(\beta)$ and $V_{100}(\beta)$ respectively, exhibit volume dependence on the parameter $\beta$. The volume of these specific nanoparticles is given by $V_1(\beta)=V_{100}(\beta)=\frac{4\pi}{3}[1250000\beta-561000] \text{nm}^3$.}
\label{Figure.3}
\end{figure*}
To explore the potential of localized states in one-dimensional radiative systems exhibiting topological features, we investigate systems characterized by an asymmetric response matrix (refer to Section.~(\ref{defect}) for the conditions under which localized states exist). Specifically, our focus lies on systems possessing mirror reflection symmetry, commonly referred to as P-symmetry. For simplicity, we assume that the volumes of particles $m$ and $n$ in the chain are dependent on $\beta$, denoted as $V_m=V_n=V(\beta)$, where $n=N+1-m$. Through the appropriate selection of a function $V(\beta)$, it becomes possible to construct an asymmetric response matrix that facilitates the description of the emergence of localized states as influenced by the coupling parameter $\beta$. In our study, we consider a simplified scenario where the volumes of both the $m$th and $n$th nanoparticles are given by $V_m=V_n=\frac{4\pi}{3}[1250000\beta-561000]\text{nm}^3$, while all other nanoparticles have constant volumes $V_{i\neq m,n}=\frac{4\pi}{3}50^3\text{nm}^3$. This choice results in a linear dependence of the particle volumes $m$ and $n$ on the intra-cell distance factor $\beta$. The specific form of $V(\beta)$, detailed in Section~(\ref{defect}) of the Appendix, is designed to maintain flat bands associated with localized states for $\beta<0.5$.
Figure~(\ref{Figure.1}c) displays the eigenvalue spectrum of the response matrix for a specific configuration $(m,n)=(1,60)$, employing the same parameters as in Figure~(\ref{Figure.1}b), except for $V_m=V_n=\frac{4\pi}{3}[1250000\beta-561000]\text{ nm}^3$. At $\beta=0.54$, the spectrum closely resembles the previous case, exhibiting a discernible gap. However, at $\beta=0.46$, a band gap also emerges, effectively dividing the spectrum into lower and upper bands. Notably, within this band gap, the presence of edge states becomes apparent. As subsequently demonstrated, these edge states exhibit localization at the boundaries of the chain and showcase resilience against perturbations. The existence of these edge states signifies the nontrivial topological nature of the system, which we will further elucidate.

The full eigenvalue spectrum of the response matrix of the system with an asymmetric response matrix is depicted in Fig.~(\ref{Figure.2}). The spectrum reveals distinct features depending on the value of the intra-cell distance factor, $\beta$. For values of $\beta$ below 0.5, localized states can be observed in the spectrum. These localized states coexist with the bulk states, giving rise to a rich and diverse eigenvalue spectrum. The \emph{bulk state} refers to the thermal modes that could mainly be excited  in the system within its bulk or bulk region, away from any surfaces, interfaces, defects, or edges. These state form allowed allowed relaxation time levels that represents the thermal properties of the system asa whole. In contrast, \emph{edge state}, \emph{localized state}. or \emph{defect states}  refers to a unique state that arises at the boundary, or defected position in a system with nontrivial topological properties. Due to their localized nature, they are often confined to a narrow region in the system. They exhibit unique properties such as unidirectional transport, robustness against disorder, and the presence of relaxation time levels within the band gap. We designate this regime as the topologically nontrivial phase (TNP). Conversely, for values of $\beta$ above 0.5, the spectrum exhibits a different behavior, devoid of distinct localized states. We refer to this regime as the topologically trivial phase (TTP). In the eigenvalue spectrum, the eigenstates are denoted by subscripts $F$ and $S$, representing the eigenstate with the largest and smallest eigenvalue, respectively. These eigenstates are commonly referred to as the fast and slow modes, respectively, based on their contributions to the decay rate and thermalization process in the chain. The presence of both bulk states and localized states in the spectrum highlights the complex interplay between the topological properties of the system and the tuning parameter $\beta$. This interplay governs the emergence of different types of eigenstates and provides insights into the system's thermal dynamics and behavior.

\section{ Profile of the Eigenstates for a Chain with asymmetric Response Matrix}\label{sec42}
The temporal evolution of temperatures, as described by Eq.(\ref{eq4}), is influenced not only by the eigenvalues but also by the eigenstates of the response matrix. In this section, we focus on examining the eigenstates of a chain of nanoparticles (NPs) with an asymmetric response matrix in its topologically nontrivial phase. To illustrate this, we consider a specific case with $\beta=0.46$, and present the profiles of the eigenstates in Fig.~(\ref{Figure.3}). The chain of NPs under investigation is the same as in the previous section, but with an extended length of $N=100$.

As described in the Appendix, the system demonstrates parity preservation, characterized by the commutation relation $[\hat \Pi,\hat H]=0$, where $\hat \Pi$ represents the parity operator. Consequently, the eigenstates can be classified as either even or odd with respect to $l/2$, where $l=x_N-x_1$. The profiles of selected modes are displayed in Fig.~(\ref{Figure.3}). We observe that the eigenstates exhibit a distinct pattern of even or odd symmetry with respect to the center of the chain. Notably, the slowest mode $\psi_1(x)$ exhibits an even-parity profile. Furthermore, the lower and upper edge states exhibit symmetric and asymmetric profiles, respectively, with both states localized at sites $i=1$ and $i=100$. As described in Sec.~(\ref{robust}) of the appendix, these modes exhibit topological robustness against perturbations that preserve the mirror reflection symmetry within the system.
\begin{figure}
\includegraphics[scale=1.1]{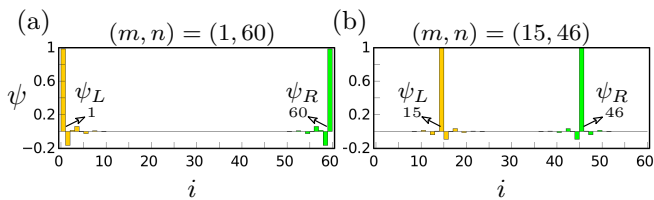}
\caption{The profiles of the L-type and R-type modes are shown for a chain of $N=60$ nanoparticles with an asymmetric response matrix in the topologically nontrivial phase of the system, where $\beta=0.46$. The specific cases depicted are (a) $(m,n)=(1,60)$ and (b) $(m,n)=(15,46)$.}
\label{Figure.4}
\end{figure}
\section{L-type and R-type eigenstates} 
Upon calculating the eigenstate profiles, we can construct L-type and R-type eigenstates as $\psi_L(x)=[\psi_D(x)-\psi_U(x)]/\sqrt{2}$ and $\psi_R(x)=[\psi_D(x)+\psi_U(x)]/\sqrt{2}$, respectively. These L-type and R-type states exhibit the characteristics of maximum localization and have a significant influence on the thermalization process, particularly in the topologically nontrivial phase of the system. Therefore, we conduct a comprehensive investigation of their properties.

In Fig.~(\ref{Figure.4}), we present the profiles of these hybridized eigenstates for a chain of $N=60$ nanoparticles, considering the special cases of $(m,n)=(1,60)$ and $(m,n)=(15,46)$. The figure depicts the profiles for a specific value of $\beta=0.46$, representing the TNP of the chain. It is evident that the L-type state is fully localized at site $m$, while the R-type state is localized at site $n$. It is important to emphasize that the L-type and R-type states are pure states, serving as simultaneous eigenstates of the response matrix in the limit of $\beta\ll \beta_c$ (for the definition of pure, mixed, and random states, refer to the Sec.~(\ref{sec3}) of the Appendix). However, since the L-type and R-type states do not possess even or odd parity, they are not parity eigenstates.
\begin{figure}
\includegraphics[scale=1]{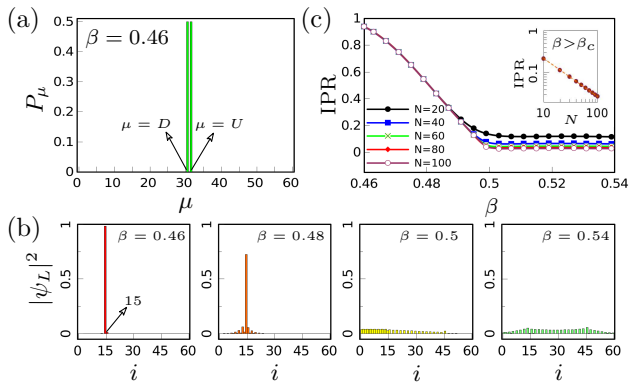}
\caption{Key features of the L-type eigenstate, specifically for the case of $(m,n)=(15,46)$, in an asymmetric SSH chain of nanoparticles are presented. (a) The weight distribution of $|\psi_L(x)|^2$ is shown for a chain of $N=60$ nanoparticles at $\beta=0.46$. (b) The probability distribution of $\psi_L(x)$ is displayed for $\beta$ values ranging from 0.46 to 0.54 in a chain with an asymmetric response matrix and $N=60$ nanoparticles. (c) The Inverse Participation Ratio (IPR) of $\psi_L(x)$ is plotted as a function of $\beta$ for chains of nanoparticles with different lengths $N$. The inset exhibits the scaling behavior of IPR vs $N$ in the topologically trivial phase (TTP)}
\label{Figure.5}
\end{figure}

In Fig.~(\ref{Figure.5}a), we illustrate the weight distribution of the L-type states in the topologically nontrivial phase for $\beta=0.46$ and $(m,n)=(15,46)$. The weight is equally distributed over the localized modes $\psi_U$ and $\psi_D$, i.e., $P_\mu=0.5[\delta_{\mu,D}+\delta_{\mu,U}]$, where $D=N/2+1$ and $U=N/2+2$. However, as we slightly increase $\beta$, both the L-type and R-type states extend over neighboring sites, no longer remaining eigenstates of the response matrix. This behavior indicates that they become delocalized. In Fig.~(\ref{Figure.5}b), we show the profile of $|\psi_L(x_i)|^2$ for different values of $\beta$. It is observed that this mode gradually extends as $\beta$ increases and eventually becomes completely delocalized over the entire chain beyond the critical point. Similar arguments apply to the localized states $\psi_U$ and $\psi_D$. For more detailed information, please refer to Sec.(\ref{LR}) of the Appendix.

The inverse participation ratio (IPR) is a valuable metric for characterizing the localization properties of a given state. In our study, where we have established the presence of edge states or localized states, we utilize the IPR defined as $\text{IPR}\mu=\sum_i |\psi_\mu(x_i)|^4/\left(\sum_i |\psi_\mu(x_i)|^2\right)^2$~\cite{Calixto2015}. Here, $\psi_\mu(x_i)$ represents the amplitude of the $\mu$th eigenstate at the $i$th nanoparticle location. The inverse of the IPR provides an estimate of the number of nanoparticles involved in the thermalization process through that particular mode. To illustrate the variation in localization quality among the eigenstates, we present the eigenstate spectrum in Fig.~(\ref{Figure.2}) using a color scale. The color bars on the right side of the figure indicate the IPR values for the eigenstates. It can be observed that most eigenstates exhibit extended behavior, except for two localized modes around $\lambda_e$ when $\beta<0.5$. Notably, as $\beta$ decreases, the localization of eigenstates in the upper localized band (UEB) and lower localized band (DEB) significantly increases.

To demonstrate the delocalization of the L-type state at the critical point, we analyze the IPR of $\psi_L(x)$ for various system sizes in Fig.~(\ref{Figure.5}c). From this plot, it can be observed that IPR$_L$ is relatively large ($\sim 1$) at $\beta=0.46$ and decreases as $\beta$ increases. The minimum value of IPR$_L$ occurs at the critical point $\beta_c=0.5$ for different system sizes $N$, indicating the transition between edge and bulk states. In the topologically trivial phase, the IPR of $\psi_L(x)$ remains approximately constant, suggesting complete delocalization of this state throughout the chain. It is evident from the figure that, in this phase, the state $\psi_L(x)$ becomes more delocalized with increasing system size. Additionally, we observed a decrease in IPR$_L$ following a scaling law of IPR$_L\sim N^{-0.9}$ with respect to the chain length $N$, as shown in the inset of Fig.~(\ref{Figure.5}c).

 \section{Dynamic of Temperatures in Topologically Trivial and Non-Trivial Phases.}  \label{sec5}

\begin{figure}
\includegraphics[scale=0.7]{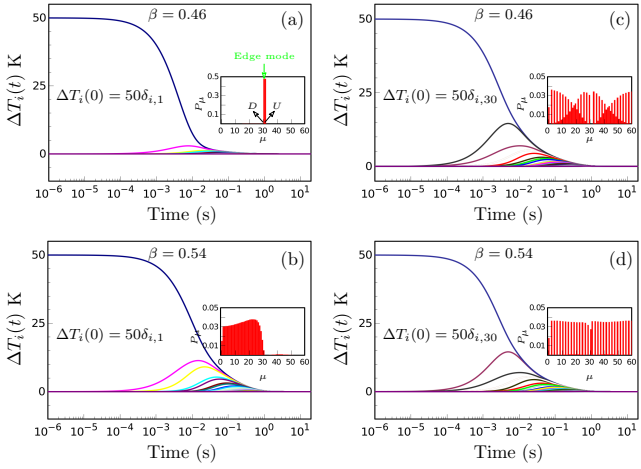}
\caption{Temperature evolution in a  chain with asymmetric response matrix comprising $N=60$ nanoparticles in both the topologically trivial and non-trivial phases. The volume of particles is determined as follows: $V_{i\neq 1,60}=\frac{4}{3}\pi \times (50)^3 \text{nm}^3$, and $V_1(\beta)=V_{60}(\beta)=\frac{4\pi}{3}\left[1250000\beta-561000\right] \text{nm}^3$. The initial temperature condition is set such that only the $j$-th particle is heated to $350$ K, while all other particles are initially at room temperature of $300$ K. It should be noted that the edge state is localized at sites 1 and 60 in the topologically non-trivial phase.
(a) Temperature evolution for $j=1$ in the topologically non-trivial phase with $\beta=0.46$.
(b) Temperature evolution for $j=1$ in the topologically trivial phase with $\beta=0.54$.
(c) Temperature evolution for $j=30$ in the topologically non-trivial phase with $\beta=0.46$.
(d) Temperature evolution for $j=30$ in the topologically trivial phase with $\beta=0.54$.}
\label{Figure.6}
\end{figure}

We are now ready to explore the dynamics of temperature in the presence of a localized state. To achieve this, we investigate the temporal evolution of temperatures by exciting edge or bulk states in a chain of $N=60$ nanoparticles with $(m,n)=(1,60)$. The initial condition for temperatures is defined such that only one of the particles (either particle 1 or particle 30) is heated up to $350$ K, denoted as $\Delta T=50$ K, while all other particles are initially at room temperature of $300$ K, denoted as $\Delta T=0$ K. The thermalization process is presented in Fig.~(\ref{Figure.6}) for both the topologically non-trivial phase (e.g., $\beta=0.46$) and the topologically trivial phase (e.g., $\beta=0.54$). The inset in the figures represents the weight distribution of the initial temperature state over modes.

The temporal evolution of temperatures in case where only particle 1 is heated up to $350$~K is shown in Figs.~(\ref{Figure.6}a) for $\beta=0.46$. Initially we may think the initial temperature state vector $\Delta T_i(0)\equiv \Delta T(x_i,0)=50\delta_{i1}$ is a mixed state. However, we observe the weight distribution of the initial temperature state is highly localized on edge state modes, i.e., $P_\mu\simeq 0.5[\delta_{\mu D}+\delta_{\mu U}]$, see the inset in Fig.~(\ref{Figure.6}a). Comparing this distribution with that in Fig.~(\ref{Figure.5}a), we can conclude that the initial temperature state is similar to $\psi_L(x)$, i.e., $\Delta T (x_i,0)\simeq 50\psi_L(x_i)$, and therefore it is semi-pure. As a result, we expect a relatively isolated thermalization (compared to the following cases) with decay rate $\sim \lambda_{e}$ in this case.

Figure (\ref{Figure.6}b) illustrates the scenario corresponding to the topologically trivial phase ($\beta=0.54$) with the same initial condition: $\Delta T_i(0)=50\delta_{i1}$. Upon examining the inset of Fig. (\ref{Figure.6}b), we observe that the distribution of $P_\mu$ extends across several modes, primarily $\mu\leq U$. Consequently, this initial temperature state can be regarded as a mixed state. In this particular case, the slowest mode $\psi_S(x)$ dominates the temporal evolution of temperatures, leading to an increased thermalization time. To further elucidate the thermalization process, Figs. (\ref{Figure.6}c) and (\ref{Figure.6}d) present the evolution when the 30th particle is heated to $350$ K initially, denoted as $\Delta T_i(0)=50\delta_{i30}$. Irrespective of the system's topological phase, the initial temperature state in this scenario is mixed, with contributions from almost all modes, as depicted in the insets of Figs. (\ref{Figure.6}c) and (\ref{Figure.6}d). Similar to the previous case, the thermalization process is primarily determined by the eigenvalue of the slowest mode, resulting in $\Delta T(x_{30},t>\lambda_S^{-1})\propto \exp(-\lambda_S t)$. Consequently, the thermalization time is significantly prolonged.
A thorough comparison of the obtained results with those derived from a system with symmetric response matrix, as detailed in Sec.~(\ref{temporalinhamiltonian}) of the Appendix, highlights an important observation. The absence of localized states in chian of identical nanoparticles yields a distinctive long-range characteristic in the propagation of heat flux along the chain. This finding implies that the presence or absence of localized states significantly impacts the transport properties and dynamics of temperatures in the system.
\section{Thermalization time}

To gain a deeper understanding of the impact of structural topology on the transient regime of temperature dynamics, we conducted calculations to determine the thermalization time of the system. Similar to the approach in the previous section, we utilized an initial temperature state to excite the desired states. For this purpose, the bulk states were excited by heating up particle $j$ in the chain, where $j\notin\{m,n\}$ in the topologically non-trivial phase. Likewise, the edge states were excited by heating up particle $j\in\{m,n\}$ in topologically non-trivial phase. In both scenarios, particle $j$ was heated to $350$ K, while all other particles were initially at room temperature of $300$ K. Subsequently, we tracked the thermalization process until the system reached its equilibrium state, known as the stable state.
To quantify the results, the thermalization time $\tau$ in each case was defined as the time at which $\Delta T(x,\tau)<10^{-12}$ K.

\begin{figure}
\includegraphics[scale=1]{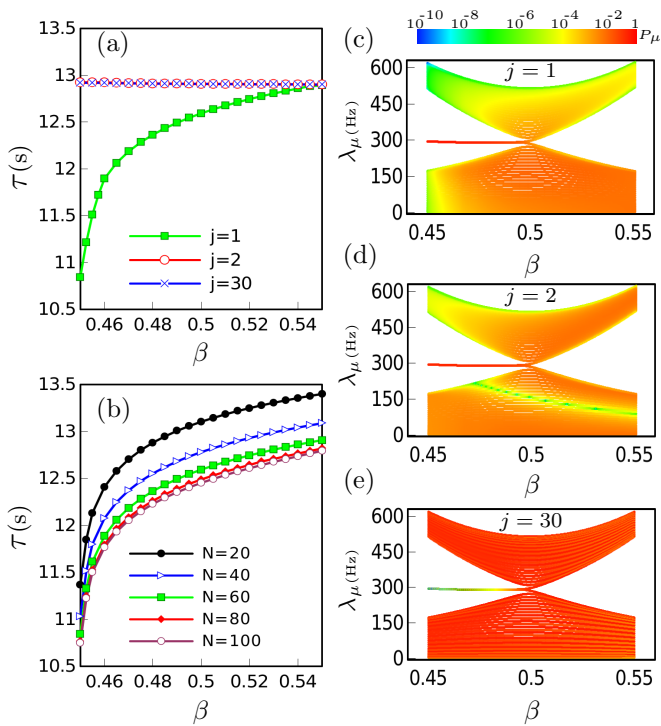}
\caption{ (a) Thermalization time of a chain with a length of $N=60$, representing the same setup as shown in Fig. (\ref{Figure.1}c), but with different initial temperature states, specifically $\Delta T(x_i,0)=50\delta_{i,j}$ for $j\in\{1,2,30\}$.
(b) Thermalization time for the initial temperature state $\Delta T(x_i,0)=50\delta_{i,1}$ in a chain of nanoparticles with different lengths, $N\in\{20,40,60,80,100\}$.
(c-e) Color map depicting the weight distribution $P_\mu$ for the initial temperature state $\Delta T(x_i,0)=50\delta_{i,j}$, where $j$ is equal to 1, 2, and 30, respectively.}
\label{Figure.7}
\end{figure}

In Fig.~(\ref{Figure.7}a), we present the thermalization time in a chain consisting of $N=60$ NPs with $(m,n)=(1,60)$ for selected values of $j$, specifically $j\in\{1,2,30\}$. The thermalization time for $j=1$ exhibits a sharp increase for $\beta\ll \beta_c$, followed by a more gradual rise in the topologically trivial phase. Interestingly, the thermalization times $\tau_2$ and $\tau_{30}$ are approximately equal and do not significantly depend on the topological phase of the system. They exhibit a slow decrease and approach the value of $\tau_1$ as $\beta\gg \beta_c$. Notably, this behavior is consistent for chains with localized states at arbitrary positions $(m,n)$.

The impact of chain length on the thermalization time is illustrated in Fig.~(\ref{Figure.7}b) for the specific case of $j=1$. Heating up particle $1$ in topologically nontrivial phase, particularly when $\beta\ll \beta_c$, excites the L-type edge state. Consequently, the thermalization time exhibits little dependence on the chain length in this limit. On the other hand, in the topologically trivial phase, we observe that the thermalization time is enhanced compared to topologically nontrivial phase and reaches a saturation point for longer chain lengths.

To gain insights into the underlying physics of the thermalization process, we examine the color maps of the weight distribution $P_\mu$ for the initial temperature state $\Delta T(x_i,0)=50\delta_{i,j}$ as a function of $\beta$. Figures (\ref{Figure.7}c)-(\ref{Figure.7}e) present these color maps, displayed in a logarithmic scale for clarity. Notably, heating up particle $1$ efficiently excites the edge state in the topologically non-trivial phase, as depicted in Figure (\ref{Figure.7}c). The weight distribution of this initial temperature state, particularly for $\beta\ll\beta_c$, exhibits localization around $\lambda_e$, similar to the behavior of $\psi_L(x)$ shown in Figure (\ref{Figure.5}a). Consequently, the initial temperature state can be regarded as semi-pure, primarily exciting the localized L-type state. As a result, we expect a rapid and nearly isolated temperature decay for the excited edge site, characterized by $\Delta T(x_{1},t)\sim 50\exp(-\lambda_e t)$, while $\Delta T(x_{i\neq 1},t)\simeq 0$. This finding excellently agrees with the results in Figs.~(\ref{Figure.7}a)-(\ref{Figure.7}b) and confirms the small values of $\tau_1$ for $\beta\ll\beta_c$. However, as $\beta$ increases beyond $\beta_c$, other modes in the bulk become populated. This is attributed to the transition of the initial temperature state from semi-pure to mixed as $\beta$ increases. Figure (\ref{Figure.7}c) demonstrates that the states in the lower bulk band are fully excited, leading to a longer thermalization time for $\beta>\beta_c$.

From Fig.~(\ref{Figure.7}d), it is evident that heating up particle $2$ results in a slight excitation of the edge states in TNP. This is expected since particle $2$ is located in close proximity to the localization site at $m=1$. Simultaneously, a significant portion of the modes in the lower bulk band is also excited. Consequently, the contribution of the slowest mode $\mu=S$ is sufficient to explain the longer thermalization time observed in TNP. In the topologically trivial phase, for any value of $\beta$, heating up particle $2$ approximately populates all bulk states in both the upper and lower bands. Once again, the thermalization time is primarily determined by mode $\mu=S$, resulting in an approximately equal value of $\tau_2$ as in TNP.

The weight distribution of the initial temperature state for $\Delta T(x_{i},0)=50\delta_{i,30}$ is considerably extended, as depicted in Fig.~(\ref{Figure.7}e). Regardless of the value of $\beta$, heating up this particle leads to excitation of all bulk states. As in the previous cases, the slowest mode dominates the thermalization time, thus suggesting a similar $\tau_2=\tau_{30}\simeq \text{constant}$ behavior, as observed in Fig.~(\ref{Figure.7}a).

 \begin{figure}
\includegraphics[scale=1]{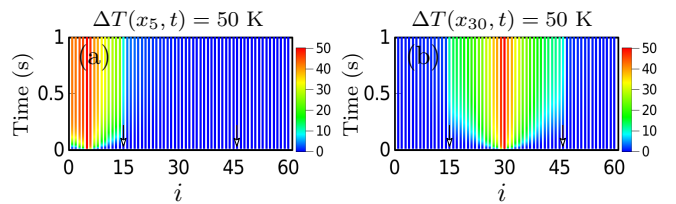}
\caption{ Spatio-temporal evolution of the temperature field $\Delta T(x_i,t)$ in a chain of $N=60$ nanoparticles for the same setup as shown in Fig.~(\ref{Figure.1}c). The localized states positioned at $(m,n)=(15,46)$ are indicated by arrows. Particle $j$ is maintained at a constant temperature $T_j=350$~K, while all other particles are initially at room temperature $300$~K. (a) $j=5$. (b) $j=30$.}
\label{Figure.8}
\end{figure}
 \begin{figure}
\includegraphics[]{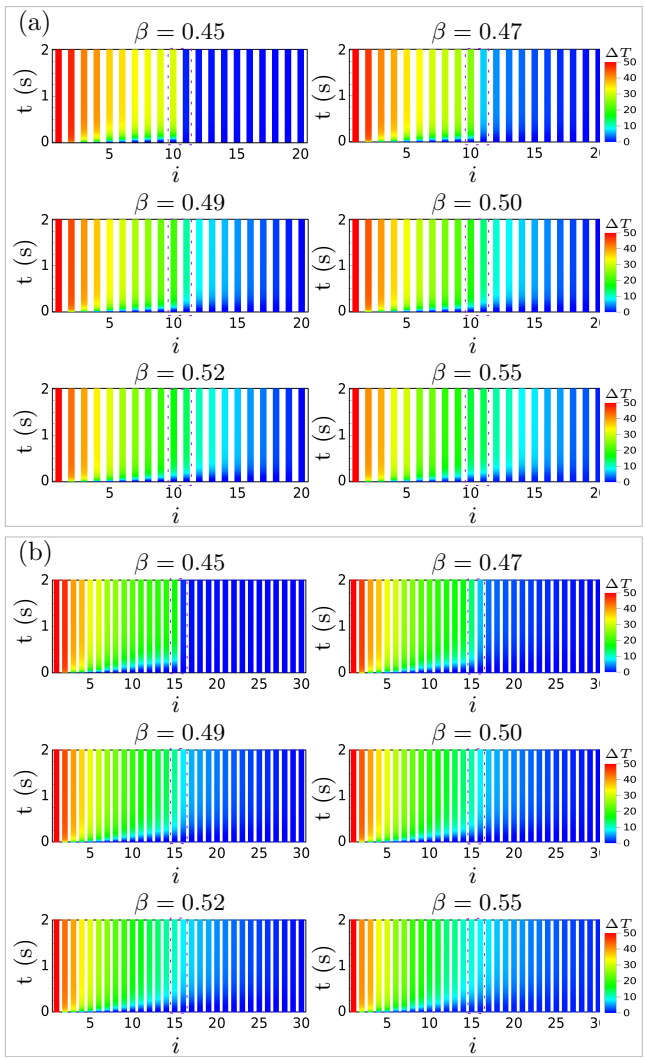}
\caption{ Spatio-temporal evolution of the temperature field $\Delta T(x_i,t)$ in a chain of $N$ nanoparticles (NPs) for the same setup as depicted in Fig.~(\ref{Figure.1}c). The positions of the localized states, indicated by dashed rectangles, are located at $(m,n)=(N/2,N/2+1)$. The first and last particles are maintained at constant temperatures $T_1=350$~K and $T_N=300$~K, respectively, while all other particles are initially at room temperature $300$~K. (a) Chain length $N=20$. (b) Chain length $N=30$.}
\label{Figure.9}
\end{figure}

\section{Radiative insulator} 
To investigate the concept of topological insulation in radiative systems, we present the spatio-temporal evolution of the temperature field $\Delta T(x,t)$ in a chain of $N=60$ nanoparticles with localized modes located at $(m,n)=(15,46)$ and $\beta=0.45$, as depicted in Fig.~(\ref{Figure.8}). In Figure.~(\ref{Figure.8}a), we fix the temperature of particle $j=5$ at $350$~K, while initially all other particles are at room temperature $300$~K. Similarly, the time evolution of the temperature field for $j=30$ is displayed in Fig.~(\ref{Figure.8}b). Since $\beta=0.45$, the system is in the topologically non-trivial phase with localized modes positioned at $x_{15}$ and $x_{46}$.

Interestingly, we observe that the localized states exhibit topological insulation behavior, effectively blocking the flow of radiative energy within the system. This leads to the formation of a sharp temperature gradient at the positions of the localized modes, with the insulation strength decreasing as $\beta$ increases. In Fig.~(\ref{Figure.8}a), where the hot source is located at $x_5$, the localized state at $x_{15}$ prevents the radiative energy from flowing across it. The same feature is observed when the hot source is positioned between the localization points of the localized states. Figure.~(\ref{Figure.8}b) clearly demonstrates that topological insulation hinders the transport of radiative energy across $x_{15}$ and $x_{46}$. It is important to note that since radiative heat transfer has a long-range nature, the insulation is not perfectly $100\%$. However, the localized modes effectively block the diffusive component of the thermal flow in the limit of $\beta\ll\beta_c$.

To further explore the influence of structure topology and the presence of localized states on the radiative heat flux, we analyze the spatio-temporal evolution of the temperature field in chains with localized states positioned at the middle of the chain, i.e., $(m,n)=(\frac{N}{2},\frac{N}{2}+1)$. Figure.~(\ref{Figure.9}) illustrates the results for chains with $N=20$ and $N=30$ NPs, showcasing the spatio-temporal evolution of the temperature field for various values of $\beta$. In both cases, the first and last particles in the chain are maintained at constant temperatures $T(x_1,t)=350$~K and $T(x_N,t)=300$~K, respectively, while all other particles initially have a temperature of $300$~K and are allowed to vary with time.

Remarkably, we observe that for $\beta=0.45$, thermal energy is not permitted to flow across the localized sites, resulting in a sharp temperature gradient at the positions of the localized modes. However, as $\beta$ increases towards the trivial topological phase, the localized modes become more extended, enabling the thermal energy to freely diffuse throughout the system. This behavior is evident in both panels (a) and (b) of Fig.~(\ref{Figure.9}), corresponding to chains with $N=20$ and $N=30$ NPs, respectively.

\section{Conclusion}

In summery, we have provided a theoretical investigation of topological phase transitions and the presence of topological modes, including edge states and conventional bulk states, in radiative systems. Using the response matrix formalism within the framework of fluctuational electrodynamics, we have explored the topological behavior of these systems. By examining the eigenvalue spectrum, we have observed distinct behaviors in systems with symmetric and asymmetric response matrices. Specifically, for systems with a symmetric response matrix, which corresponds to identical nanoparticles, we have identified topological phase transitions. However, we have not observed the existence of edge states in this case.

On the other hand, by considering systems with an asymmetric response matrix and mirror reflection symmetry (P-symmetry), we have discovered the emergence of localized states within the band gap for certain parameter values. This finding indicates the presence of a topologically nontrivial phase in the system. These localized states, known as edge states or defect modes, are confined to the boundaries and demonstrate robustness against perturbations that maintain the mirror reflection symmetry of the system.

Additionally, we examined the temporal evolution of temperatures in the presence of a localized state by exciting either edge or bulk states in a chain of nanoparticles with an asymmetric response matrix. Our observations of temperature dynamics for the topologically nontrivial and topologically trivial phases highlighted the influence of the system's topology on thermal behavior.

Furthermore, our investigation of the spatio-temporal evolution of the temperature field in the presence of localized states revealed the topological insulation characteristics of these states. The localized states effectively blocked the flow of radiative energy within the system, resulting in the formation of sharp temperature gradients at their positions.

Overall, our research enhances the understanding of heat transfer in radiative systems from a topological perspective. It provides valuable insights into topological phase transitions, the existence of localized states, and the impact of topology on the thermalization process. The approach we propose has the potential to inspire future research in nanoparticle ensembles, both in two-dimensional and three-dimensional settings, as well as in structures with planar geometry that rely on radiation for energy transfer. Furthermore, our approach can be extended to investigate the active control of topological features using external electric or magnetic fields. We believe that our findings reveal new and fascinating aspects of topological phases in radiative systems and offer valuable insights into the topological insulation of thermal energy, which may find applications in the field of radiative heat transfer.


\renewcommand{\theequation}{A.\arabic{equation}}
\renewcommand{\thefigure}{A.\arabic{figure}}
\setcounter{equation}{0} 
\setcounter{figure}{0} 
\renewcommand{\thesection}{A.\arabic{section}}
\setcounter{section}{0} 
\section*{APPENDIX} 

In this note, we present a scientific formulation of temperature dynamics in a system of particles interacting through radiation. Our approach successfully describes the observation of edge states and topological phase transitions in many-body systems. The calculations are conducted within the theoretical framework of fluctuational electrodynamics, specifically employing the dipolar approximation.
By applying linear response theory, we introduce a theoretical framework to analyze the temporal evolution of temperatures in a system of spherical nanoparticles (NPs). This formalism allows us to describe the dynamics in terms of eigenstates of the response matrix. Consequently, we can classify the system's dynamical properties based on their initial temperature state. To demonstrate the occurrence of topological phase transitions, we utilize the introduced eigenmode representation in a chain of particles with an asymmetric response matrix. Additionally, we explore techniques for manipulating the spatial position of localized states within the chain.

To investigate the localization of modes and the impact of chain size on the localization of topological edge states, we employ the inverse participation ratio (IPR) as a useful tool. Finally, we analyze the thermalization process in the NPs chain and compare the thermalization time between topologically trivial and non-trivial phases of the system. Our formalism establishes a strong foundation for exploring the topology and related phenomena in radiative systems.
\section{Theory and Model}\label{sec2}
\begin{figure}
\includegraphics[]{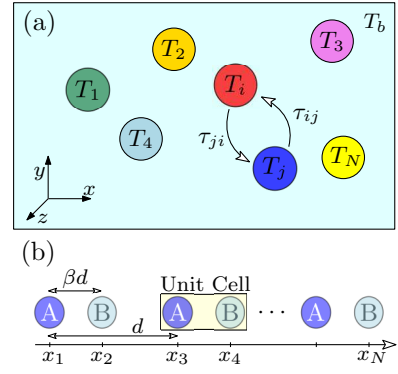}
\caption{(a) Sketch of a radiative system of $N$ spherical NPs with initial temperature $T_i(0)~(i=1,2,\cdots,N)$ immersed in a thermal bath at $T_b$. (B) A schematic illustration of bipartite chain of spherical NPs along the $x$ axis with lattice constant $d$ and immersed in a thermal bath at $T_b=300$~K. The volume of particles is $V_i(\beta),~(i=1,2,\cdots,N)$ and the intra-cell separation distance  between NPs A and B is $\beta d$.}
\label{FIGA1}
\end{figure}

Let us consider a system comprising $N$ spherical nanoparticles, each characterized by its volume $V_i$, volumetric specific heat capacity $c_i$, and initial temperatures $T_i(0)$. These nanoparticles are immersed in a thermal bath maintained at a temperature of $T_b = 300$ K, as depicted in Fig.~(\ref{FIGA1}a). Following previous studies, the evolution of nanoparticle temperatures can be described by a set of coupled differential equations given as follows:

\begin{equation}
\gamma_i\frac{d T_i}{d t}=\mathcal{P}_i ~~~~(i=1,2,\cdots,N).
\label{eqs1}
\end{equation}
Here, $\gamma_i = c_iV_i$ represents the heat capacity of the $i$-th nanoparticle, and $\mathcal{P}_i$ denotes the total power dissipated in that particular nanoparticle. To calculate $\mathcal{P}_i$, we utilize the fluctuation-dissipation theorem and the dipole approximation, yielding:
\begin{equation}
\mathcal{P}_i=\int_0^\infty \frac{d\omega}{2\pi}\sum_{j=1}^N\Big[\tau_{ij}\Theta (\omega,T_j)- \tau_{ij}\Theta (\omega,T_b)\Big].
\label{eqs2}
\end{equation}

In linear approximation, we can expand $\Theta(\omega, T_j)$ in Eq.~(\ref{eqs2}) about $T_b$ by writing $T_j=T_b+\Delta T_j$ to get~\cite{PhysRevLett126193601,PhysRevB104024301,DONG2022123318}
\begin{equation}
\mathcal{P}_i=\sum_{j=1}^NF_{ij}\Delta T_j,
\label{eqs3}
\end{equation}
where 
\begin{equation}
F_{ij}=\int_0^\infty \frac{d\omega}{2\pi} \tau_{ij}(\omega)\frac{\partial \Theta(\omega,T)}{\partial T}{\Big |}_{T_b}.
\label{eqs4}
\end{equation}
By substituting Eq.~(\ref{eqs3}) into Eq.~(\ref{eqs1}), we get
\begin{equation}
\frac{d}{dt}\Delta{  T}_i(t)=\gamma_i^{-1} \sum_{i=1}^NF_{ij}\Delta T_j.
\label{eqs5}
\end{equation}
Now, by defining temperature state vector $\Delta{\mathbf{T}}=(\Delta T_1,\Delta T_2,\cdots,\Delta T_N)^\intercal$, we can write Eq.~(\ref{eqs5}) as
\begin{equation}
\frac{d}{dt}\Delta{ {\bf T}}(t)=-{\hat H}\Delta{ {\bf T}}(t).
\label{eqs6}
\end{equation}
In the provided equation, the linear response matrix is denoted as $\hat H = -\hat\Gamma^{-1}\hat F$ ~\cite{PhysRevB104024301}, where $\hat\Gamma$ represents a diagonal matrix with heat capacities along the diagonal, given by $\hat\Gamma = \text{diag}\{\gamma_1,\gamma_2,\cdots,\gamma_N\}$. In a system where all particles are of the same material, i.e., $c_1=c_2=\cdots=c_N\equiv c$, we can express $\hat{\Gamma}$ as $c~\text{diag}\{V_1,V_2,\cdots,V_N\}$. Furthermore, the conductance matrix $\hat F$ is always real and symmetric, irrespective of whether the particles are identical or not.

Notice the similarity in structure between Equation (\ref{eqs6}) and the Schrödinger equation in quantum mechanics, given by $|\dot\psi\rangle=-i\hbar^{-1}\hat H|\psi\rangle$. To establish the Hamiltonian for dissipative systems, we can define $\hat H'=-i\hat H$ with purely imaginary eigenvalues. However, for simplicity and without loss of generality, we present the theory in terms of the response matrix with real eigenvalues.

In the case of identical particles, where $\hat\Gamma$ is diagonal and $\hat F$ is a symmetric matrix, the response matrix $\hat H$ is symmetric. However, in general, the heat capacity of particles may differ ($\gamma_m\neq\gamma_n$), especially when the volumes ($V_m\neq V_n$) are unequal. This non-identical nature renders the response matrix asymmetric, represented as $[H,H^\dag]\neq 0$.

Our approach involves determining the eigenvalues and eigenvectors of the response matrix $\hat H$ and constructing the propagator $\hat U(t)$ using these eigenvalues and eigenvectors. Since the response matrix is time-independent, this step is relatively straightforward. After obtaining $\hat U(t)$, we express the state vector $\Delta{ {\bf T}}(t)$ as $\hat U(t)\Delta{ {\bf T}}(0)$. To proceed, we expand the vector representing the initial temperature state using the set of eigenvectors $\boldsymbol{\psi}\mu$:
\begin{equation}
\Delta{ {\bf T}}(0)=\sum_{\mu=1}^N\boldsymbol{\psi}_\mu[\boldsymbol{\phi}_\mu^\intercal\cdot \Delta{ {\bf T}}(0)],
\label{eqs7}
\end{equation}
Here, "$\cdot$" denotes the inner product, and we assume the linear independence of the eigenvectors $\boldsymbol{\psi}_\mu$ while considering that the corresponding left eigenvectors $\boldsymbol{\phi}_\mu$ are normalized such that $\boldsymbol{\phi}_\mu^\intercal\cdot\boldsymbol{\psi}_\mu=1$. In our formalism, $C_\mu(0)=\boldsymbol{\phi}_\mu^\intercal\cdot\Delta{ {\bf T}}(0)$ represents the weight of the initial temperature state in the $\mu$th mode. Expanding $\Delta{ {\bf T}}(t)$ in a similar manner:
\begin{equation}
\Delta{ {\bf T}}(t)=\sum_{\mu=1}^NC_\mu(t)\boldsymbol{\psi}_\mu,
\label{eqs8}
\end{equation}
By substituting Eqs.(\ref{eqs7}) and (\ref{eqs8}) into Eq.(\ref{eqs6}), we can obtain the expression for $\hat U(t)$ as follows:
\begin{equation}
\hat U(t)=\sum_{\mu=1}^N\text{e}^{-\lambda_\mu t}\boldsymbol{\psi}_\mu \boldsymbol{\phi}_\mu^\intercal .
\label{eqs9}
\end{equation}
Therefore, the time evolution of the temperature  state is given by:
\begin{equation}
\Delta{ {\bf T}}(t)=\hat U(t)\Delta{ {\bf T}}(0)=\sum_{\mu=1}^NC_\mu(0)e^{-\lambda_\mu t}\boldsymbol{\psi}_\mu,
\label{eqs10}
\end{equation}
which leads to the equation:
\begin{equation}
C_\mu(t)=C_\mu(0)e^{-\lambda_\mu t}.
\label{eqs11}
\end{equation}
The temporal behavior of temperatures can be effectively described using continuous notation. Let us consider a linear chain of nanoparticles positioned along the $x$-axis, with a total length of $l=x_N-x_1$. The one-dimensional temperature field, denoted as $\Delta T(x,t)$, represents the variation in temperature within the chain relative to the thermal bath, given by $\Delta T(x,t)=T(x,t)-T_b$. Thus, the evolution of the temperature field over time can be expressed as:

\begin{equation}
\Delta T(x,t)=\sum_{\mu=1}^{N} C_\mu(0)e^{-\lambda_\mu t} \psi_\mu(x),
\label{eqs12}
\end{equation}

In this equation, $\psi_\mu(x)$ corresponds to the right eigenstate of the response matrix associated with the eigenvalue $\lambda_\mu$. The probability density of mode $\mu$ is defined as $\rho_\mu(x)=|\psi_\mu(x)|^2$. Additionally, we introduce the inverse participation ratio (IPR), which is defined as:

\begin{equation}
\text{IPR}\mu=\frac{\sum{i=1}^N |\psi_\mu(x_i)|^4}{\left[\sum_{i=1}^N |\psi_\mu(x_i)|^2\right]^2},
\label{eqs13}
\end{equation}

to quantify the spatial localization of the eigenstates. It is worth noting that the IPR of the eigenstates lies within the range of $[1/N,0.5]$, where the lower limit corresponds to a completely extended mode and the upper limit corresponds to a completely localized mode. As we will demonstrate, the first eigenmode $\psi_1(x)$ is completely extended, resulting in $\text{IPR}_1=1/N$. Moreover, if we construct a combination of eigenstates, the IPR of the constructed state will fall within the range of $[1/N,1]$.

To investigate the topological phase transition in radiative systems, we examine a bipartite chain of spherical nanoparticles  arranged along the $x$-axis, with volumes $V_i$ for $i=1,2,\cdots,N$. These NPs are immersed in an external thermal bath at a temperature of $T_b=300$ K. As illustrated in Figure (\ref{FIGA1}b), the separation distance between cells remains constant at $d=500$ nm, while the intra-cell distance factor $\beta$ is employed to manipulate the system's topological properties. We begin by emphasizing key aspects of the introduced formalism, which aid in comprehending the temporal evolution of temperatures in the topological edge modes.

\section{Temporal Evolution of temperatures in the presence of constraint.}\label{constatntT}
In the previous section, we examined the temporal evolution of temperature in the absence of external power, where the temperature in the phase space always reached a steady state of $\Delta T_i(t\to\infty)=0$, regardless of the initial temperature state. Now, let us consider a scenario where we want to maintain the temperature of certain particles at a constant value and study the temporal evolution of the temperatures of the remaining particles. Therefore, Equation (\ref{eq1}) can be expressed in a more general form as:
\begin{equation}
\gamma_i\frac{d T_i}{d t}=\mathcal P_i+\mathcal F_i^e(t),~~(i=1,2,\cdots,N),
\label{eqn1}
\end{equation}
where $F_i^e(t)$ is in general a time-dependent external power applied to particle $i$ to keep its temperature fixed. For sake of simplicity, we assume that the temperature of $p$th particle is taken to be fixed at $T_p(t)=T_p(0)$. Equation.~(\ref{eqn1}) then can be written as
\begin{equation}
\gamma_i\frac{d T_i}{d t}=\mathcal P_i+\mathcal F_p^e(t)\delta_{i,p},~~(i=1,2,\cdots,N),
\label{eqn1}
\end{equation}
The functionality of $F_p^e(t)$ is not known at this stage but it certainly depend on time. 
In order to calculate the temporal evolution of temperatures as well as the functionality$F_p^e(t)$, we develop a formalism in this section that can simply extend for case where more than one object's temperature is kept fixed. The number of dynamical variables in our system, specifically temperatures, is reduced to $N-1$. This reduction is achieved by defining a rearranged temperature state denoted as $\Delta{\mathbf{T}'}=({\mathbf{T}}-\mathbf{T_b})^\intercal=(\Delta T_1',\Delta T_2',\cdots,\Delta T_{N-1}')^\intercal$, where $\Delta T_p'$ is excluded. By making this adjustment, Eq.~(\ref{eqs6}) can be expressed as follows:
\begin{equation}
\frac{d}{dt}\Delta{ {\bf T}'}(t)=-{\hat H'}\Delta{ {\bf T}'}(t)+{\bf F},
\label{eqn2}
\end{equation}
In this equation, $\hat H'$ ia an $(N-1)\times(N-1)$ dimensional matrix that represents the reduced response matrix obtained by eliminating the $p$th row and $p$th column of the original matrix $\hat H$. The matrix $\hat H'$ can be represented in a biorthogonal form as $
\hat H' = \sum_{\mu=1}^{N-1} \lambda_\mu' \boldsymbol{\psi}_\mu' \boldsymbol{\phi}_\mu'^\intercal$, where $\lambda_\mu'$ are the eigenvalues of $\hat H'$, and $\boldsymbol{\psi}_\mu'$ and $\boldsymbol{\phi}_\mu'$ are the corresponding right and left eigenvectors, respectively. The eigenvectors $\boldsymbol{\psi}_\mu'$ and $\boldsymbol{\phi}_\mu'$ play a crucial role in the analysis of the system dynamics. They are orthogonal to each other and provide insight into the specific modes of behavior in the reduced system. The eigenvalues $\lambda_\mu'$ represent the associated frequencies or rates of change for each mode.
Furthermore, in the given system, the power received by particles from particle $p$ can be mathematically expressed as $F_i = H_{pi}\Delta T'_p(0)$, where $\Delta T'_p(0)$ represents the  temperature difference between $T_p(0)$ and $T_b$.

To obtain the general form of Eq.(\ref{eqs12}) while considering the maintained constraint, we expand $\Delta \mathbf{T}'(t)$, $\Delta \mathbf{T}'(0)$, and $\mathbf{F}$ using the basis vectors of $\hat H'$. By utilizing Eq.(\ref{eqn2}), the resulting equation is as follows:

\begin{equation}
\Delta T_i'(t) = \sum_{\mu=1}^{N-1} \left[ \frac{C_\mu^F}{\lambda_\mu'} + \left( C_\mu'(0) - \frac{C_\mu^F}{\lambda_\mu'} \right) e^{-\lambda_\mu' t} \right] \psi_\mu'(x_i),
\label{eqn3}
\end{equation}

In Equation (\ref{eqn3}), $C_\mu'(0)$ represents the initial weight of the temperature state in the $\mu$th mode, given by $C_\mu'(0) = \boldsymbol{\phi}_\mu'^\intercal \cdot \Delta{\mathbf{T}'}(0)$. The quantity $C\mu^F$ is determined as $C_\mu^F=\boldsymbol{\phi}_\mu'^\intercal\cdot {\bf F}$. By computing the values of $\Delta T_i(t)$, the power required to maintain the temperatures of particle $p$ fixed can be obtained using the unused equation in Eq.~(\ref{eqn1}) as $F_p^e(t)=\sum_{i=1}^NH_{pi}\Delta T_i'(t)$.
\section{Temporal Evolution: Classification of The Initial Temperature state vector.}\label{sec3}

To demonstrate the applicability of our developed formalism, we consider a specific configuration of nanoparticles as an illustrative example. However, it is important to note that the conclusions drawn are general and can be applied to any ensemble of NPs. We examine a periodic array consisting of $N=60$ identical SiC NPs with a radius of $R=50$ nm, inter-cell separation distance of $d=500$ nm, and an intra-cell distance factor of $\beta=0.5$. By utilizing Eq.~(\ref{eqs12}), we can obtain the spatio-temporal evolution of the temperature field $\Delta {T}(x,t)$ given an initial distribution $\Delta {T}(x,0)$. Through this analysis, we reveal that the dynamic behavior of the radiative system is highly dependent on the choice of initial temperature values. 
\subsection{Pure state}\label{sec31}
A particular case of interest is a "pure" state, where the initial temperature field coincides with one of the eigenstates {$\psi_\mu(x)$}. In other words, if we initialize the system with $\Delta {T}(x,0)=\psi_\nu(x)$, we obtain $C_\mu(0)=\delta_{\mu \nu}$, and at a later time, $\Delta {T}(x,t)=\psi_\nu(x)e^{-\lambda_\nu t}$. Consequently, temperatures in an initially pure state exhibit exponential decay with a rate determined by $\lambda_\nu$.

In this scenario, the particles are decoupled, and their temperatures evolve independently. As a result, inter-particle thermalization does not occur, and all particles eventually reach thermal equilibrium with the external bath simultaneously. The time required for the system to reach equilibrium in this case is highly sensitive to the specific initial pure state. Therefore, a system initially in the "fastest" state ($\psi_F=\psi_{N}$ with $\lambda_F=\max\{\lambda_\mu\}$) achieves thermalization in the shortest amount of time, while the "slowest" state ($\psi_S=\psi_1$ with $\lambda_S=\min\{\lambda_\mu\}$) experiences the longest thermalization time. 

\begin{figure*}[]
\includegraphics[]{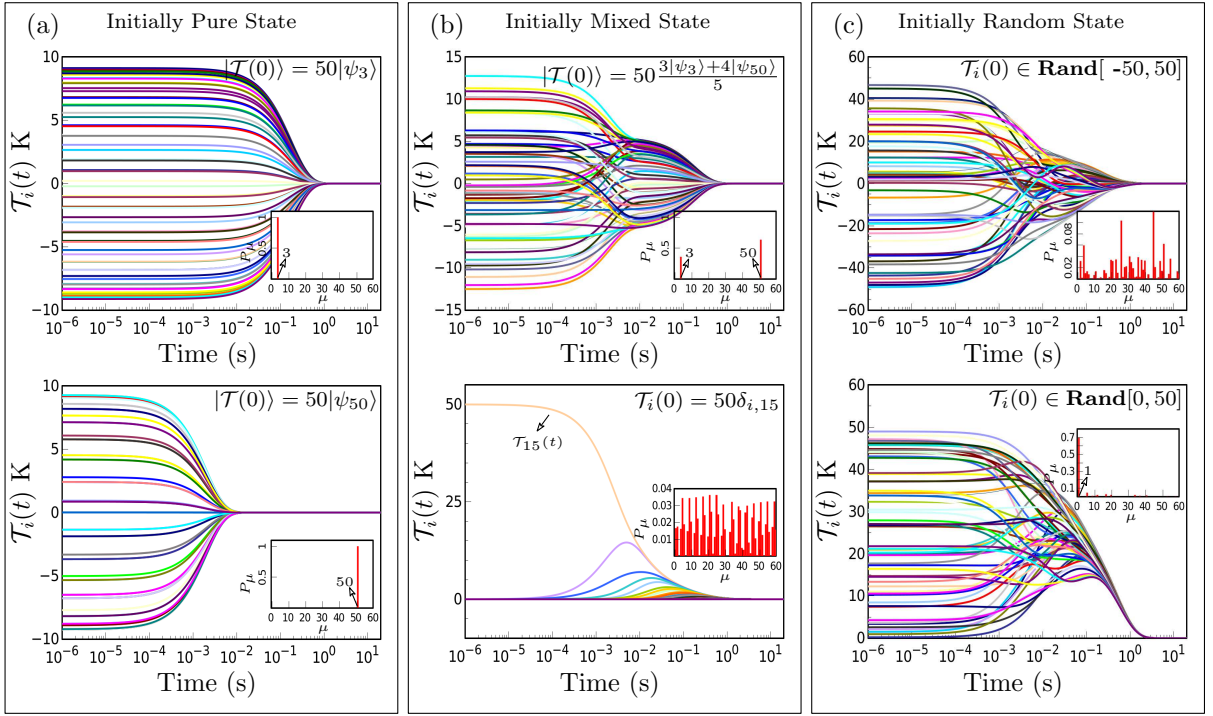}
\caption{Dynamics of temperature in a periodic array of $N=60$ identical NPs with $R=50$~nm, $d=500$~nm, and $\beta=0.5$. The insets show the  weight distribution of the initial temperature state vector. Panel.~(a) Pure initial state: In the first row $\Delta T(x,0)=50\times\psi_3(x)$ and in the second row $\Delta T(x,0)=50\times\psi_{50}(x)$. Panel.~(b) Mixed initial state: In the first row $\Delta T(x,0)=50\times[\frac{3}{5}\psi_3(x)+\frac{4}{5}\psi_{50}]$ and in the second row $\Delta T(x,0)=50\times \delta(x-x_{15})$. Panel.~(c) Random initial state: In the first row $\Delta T(x,0)\in \text {Rand}[-50,50]$ and in the second row $\Delta T(x,0)\in \text {Rand}[0,50]$. }
\label{FIGA2}
\end{figure*}

As an illustrative example, we present the temperature dynamics for two distinct pure cases in panel (a) of Fig.~(\ref{FIGA2}). In the first case, we consider $\Delta {T}(x,0)=50\times\psi_3(x)$, while in the second case, we set $\Delta {T}(x,0)=50\times\psi_{50}(x)$. It can be observed that both cases exhibit temporal temperature profiles characterized by pure exponential decay. The decay rate in each case corresponds to the eigenvalue of the respective initial pure state, namely $\lambda_3$ and $\lambda_{50}$. The insets of these figures display the weight distribution of the initial states. An interesting characteristic of the pure states is that $P_\mu$ takes the value of $\delta_{\mu 3}$ and $\delta_{\mu 50}$ for the first and second case, respectively.
\subsection{Mixed state}\label{sec32}
The initial state can be a mixed state representing an arbitrary initial temperature distribution. In this case, the system is described by a probabilistic mixture of pure states, and the combination of modes contributes to the thermalization process with an external thermal bath. The time evolution of the temperature state in this scenario involves multiple exponentially decaying terms, indicating several inter-particle thermalization events prior to complete thermalization with the bath. Consequently, the thermalization process exhibits multiple characteristic times, with the thermalization time primarily determined by the weight of the mode with the lowest eigenvalue.

As a representative example, we present in panel (b) of Fig.~(\ref{FIGA2}) the temperature dynamics for two different types of mixed states. In the first case, the initial temperature state is expressed as a linear combination of two pure states, specifically $\Delta {T}(x,0)=50\times\left(\frac{3}{5}\psi_3(x)+\frac{4}{5}\psi_{50}(x)\right)$. We observe that the system exhibits an energy relaxation timescale before thermalizing with the environment. This behavior is absent for initial states with even symmetry, such as $\Delta {T}(x,0)=50\times\left(\frac{3}{5}\psi_3(x)+\frac{4}{5}\psi_{51}(x)\right)$, or odd symmetry, such as $\Delta {T}(x,0)=50\times\left(\frac{3}{5}\psi_4(x)+\frac{4}{5}\psi_{50}(x)\right)$, due to parity conservation. Consequently, we conclude that the temperature distribution of the chain remains symmetric (asymmetric) for initially symmetric (asymmetric) temperature profiles. As depicted in the inset, the weight distribution of the initial state shows a combination weighted on modes $3$ and $50$, i.e., $P_{\mu}=\left|\frac{3}{5}\right|^2\delta_{\mu 3}+\left|\frac{4}{5}\right|^2\delta_{\mu 50}$.

Another interesting case is when only the $15$th particle is initially heated to $350$ K, while all other particles start at room temperature $300$ K, i.e., $\Delta {T}(x_i,0)=50\times \delta_{i,15}$. This case does not possess even or odd symmetry. As shown in the second row of panel (b) in Fig.~(\ref{FIGA2}), {\it all} eigenstates are initially populated. Moreover, the system undergoes $N-1=50$ inter-particle thermalization events before reaching the equilibrium state.
\subsection{Random state} \label{sec33}
Finally, we consider the case of a random distribution for the initial temperatures (referred to as a "random" state), where all modes are expected to contribute to the temporal evolution of temperatures in the system. Despite the high coupling between particles in this case, the thermalization time is primarily determined by the slowest mode $\mu=1$. Similar to the mixed case, the thermalization process may involve multiple inter-particle thermalization events before the system reaches its equilibrium state.

As a representative example within this scenario, we investigate two cases and present the results in panel (c) of Fig.~(\ref{FIGA2}). In the first case, the initial temperatures are randomly distributed within the range $\Delta {T}(x,0)\in \text{Rand}[-50,50]$, while in the second case, they are randomly distributed within the range $\Delta {T}(x,0)\in \text{Rand}[0,50]$. The main difference between these cases lies in the average initial temperature. In the first case, the average initial temperature is $\bar{\Delta {T}}(0)=0$ K, while in the second case, it is $\bar{\Delta {T}}(0)=25$ K. For the first case, the initial temperature state is highly mixed, and most of the eigenstates are initially populated. However, the non-zero average of the initial temperature distribution in the second case causes the weight distribution to be highly concentrated on mode $\psi_1$. As we will demonstrate in the next section, the first mode is constant, i.e., $\psi_1=\text{const}$. Therefore, this result is reasonable, as a non-zero average for the initial temperature state leads to a semi-pure state that predominantly populates mode $1$.

Equipped with the appropriate formalism described above, we are now ready to investigate the topological behaviors of the radiative system. The fundamental question is: under what circumstances does a topological phase transition occur in a chain of nanoparticles?

\section{Existence of The Edge State}\label{sec4}
The presence of spectrally fixed edge states at the midgap value of the spectrum requires the presence of parity symmetry. In the case of identical particles, we have $\hat{\Gamma} = \gamma\mathbbm{1}$, which implies that both $\hat{\Gamma}^{-1}$ and $\hat{F}$ exhibit even parity. As a result, the response matrix $\hat{H}$ remains invariant under parity. Consequently, we can observe that $\hat{H}\hat{\Pi} = +\hat{\Pi}\hat{H}$, and it is evident that $[\hat{H},\hat{\Pi}] = 0$, where $\hat{H} = \hat{H}^{\dagger}$. Here, $\hat{\Pi}$ represents the parity operator relative to the middle of the chain. What happens if the particles are not identical? In general, the response matrix will be asymmetric, $\hat H\neq\hat H^\dagger$. However, with a properly designed structure, there is a possibility for the system to preserve parity symmetry. Once again, $\hat F$ remains invariant under parity. However, for $\hat\Gamma^{-1}$ and, consequently, the response matrix $\hat H$ to be even under parity, we must satisfy the condition:
\begin{equation}
\begin{aligned}
\gamma_i=\gamma_{N+1-i}.
\end{aligned}
\label{eqs15}
\end{equation}
On the other hand, there must be a mirror reflection symmetry with respect to the middle of the chain. Thus, $\hat\Gamma$ will be symmetric (with respect to its minor diagonal), leading to $H_{ii}=H_{N+1-i,N+1-i}$. This condition is trivially satisfied for symmetric response matrices. Since $\gamma_i=c_iV_i$, in the special case of particles with the same material, we must have:
\begin{equation}
\begin{aligned}
V_i=V_{N+1-i}.
\end{aligned}
\label{eqs16}
\end{equation}

It is important to emphasize that even in this case, the eigenvalues are real and positive. Moreover, $\boldsymbol{\psi}_\mu$ are simultaneous eigenkets of $\hat H$ and $\hat\Pi$, just like in the symmetric case. This situation also applies to the expected edge states $\boldsymbol{\psi}_D$ and $\boldsymbol{\psi}_U$, where the indices $D$ and $U$ refer to the localized band (as discussed in the next section). As we will see, $\boldsymbol{\psi}_D$ and $\boldsymbol{\psi}_U$ are orthogonal, and we can affirm with certainty that one of them is even while the other is odd.

In order for the system to exhibit topological edge states, we require the response matrix to be degenerate, in addition to possessing parity symmetry. Furthermore, to achieve a flat spectrum of edge states in the topologically non-trivial phase, the degenerate eigenvalues must remain constant within a certain range of $\beta$. The first condition necessitates the existence of at least two states that are simultaneous eigenvectors of $\hat H$ with the same eigenvalue $\lambda_e$, namely $\hat H\boldsymbol{\psi}_D= \lambda_D\boldsymbol{\psi}_D$ and $\hat H\boldsymbol{\psi}_U= \lambda_U\boldsymbol{\psi}_U$, where $\lambda_D=\lambda_U=\lambda_e$. The latter condition establishes the presence of an edge in the system, provided that $d\lambda_e/d\beta\simeq 0$ within the desired range of $\beta$ (referred to as the "flatness rule").

Considering localized modes that are localized at lattice sites $m$ and $n$, with the special case of $(m,n)=(1,N)$ representing the edge state, we can utilize mirror reflection symmetry to derive the following relationships: $n=N+1-m$, $\gamma_m=\gamma_n$, $H_{mm}=H_{nn}$. As a result, the weight distribution of the edge states is expected to be localized on both of these sites, i.e., $|\psi_D(x)|^2=|\psi_U(x)|^2\simeq[\delta(x-x_m)+\delta(x-x_n)]/2$.

In an extremely dilute system, the response matrix takes on a diagonal form, with diagonal elements $H_{ii}\propto V_i/\gamma_i$. This accounts for the interaction with the thermal bath but does not include the effects of inter-particle interactions. For a chain of particles composed of the same material, we can express $H_{ii}\propto c_i^{-1}\equiv A$, where $A$ is a positive constant. The diagonal structure of $\hat{H}$ implies that $\lambda_e=H_{mm}=H_{nn}= A$. However, beyond the dilute limit of particles, the interactions between cells and the inter-particle interactions (many-body effects) become significant, leading to deviations from the diagonal form of the response matrix. Nonetheless, in the diagonal representation, we can approximate the eigenvalues of $\hat{H}$ as $\lambda_e\approx A+f_m(\beta,V_1,V_2,\cdots,V_N)/V_m=A+f_n(\beta,V_1,V_2,\cdots,V_N)/V_n$, where $f(\beta,V_1,V_2,\cdots,V_N)$ incorporates the interactions among the particles. Since $f_m=f_n\equiv f_e$ and $V_m=V_n\equiv V_e$, we can write $\lambda_e\approx A+f_e/V_e$. Therefore, assuming that $U$ and $D$ are localized modes located on both sites $m$ and $n$ within the desired range of $\beta$ (e.g., below the critical point $\beta<\beta_c$), the degeneracy and flatness rules discussed above imply that we must have

\begin{subnumcases}{\label{eqs17}}
   \lambda_e \simeq  A+f_e(\beta,{\bf V})/V_e\simeq \text{cte},\label{eqs17a}
   \\
   \frac{d\lambda_e}{d\beta}\simeq 0\Downarrow \label{eqs17b}\\ \nonumber\Big[ \big(\frac{df_e}{d\beta}+\sum_i\frac{\partial f_e}{\partial V_i}\frac{dV_i}{d\beta}\big)V_e-\frac{dV_e}{d\beta}f_e\Big]/V_e^2\simeq 0.
\end{subnumcases}

\subsection{ symmetric Vs asymmetric Response Matrix}\label{sec41}
Let us now consider a specific scenario involving a chain composed of identical particles, where the volume of each particle, denoted as $V_i$, remains constant throughout the chain. We will refer to this constant volume as $V_0$. As discussed earlier, in this case, the response matrix is symmetric. We can simplify Eq.~(\ref{eqs17a}) to $\lambda_e\simeq g_e(\beta)$, where $g_e(\beta)=A+f_e(\beta)/V_0$. The condition for the flatness of the edge band in a topologically non-trivial phase can be reduced to $\frac{df_e}{d\beta}\simeq 0$.

\begin{figure}[t]
\includegraphics[]{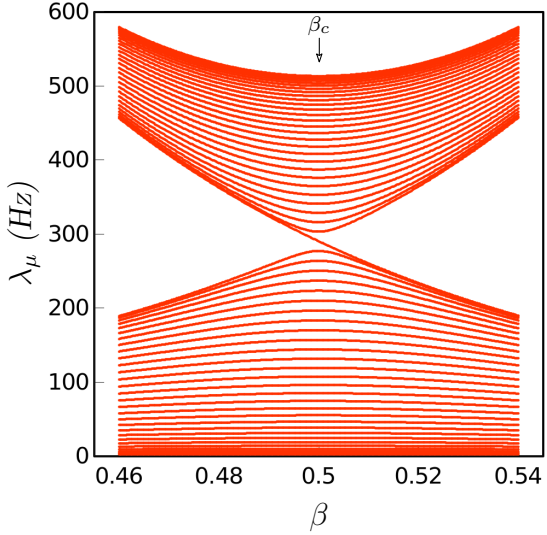}
\caption{Eigenvalue spectrum of a chain composed of SiC nanoparticles with a lattice constant of $d = 500$ nm and a total of $N = 60$ nanoparticles. The system characterized by a symmetric response matrix $\hat{H}$, where $V_1 = V_2 = \cdots = V_N = V_0$ with $V_0 = \frac{4}{3}\pi 50^3 \text{ nm}^3$.}
\label{FIGA3}
\end{figure}

However, as depicted in Fig.~(\ref{FIGA3}), we find that this equation holds true only at the critical point $\beta_c=0.5$. In the figure, we present the eigenvalue spectrum of a chain composed of $N=60$ SiC nanoparticles, all having the same volume $V_0=\frac{4}{3}\pi (50)^3\text{nm}^3$ and a lattice constant $d=500\text{nm}$. It is evident that no edge states can be found in this case, and the flatness condition is satisfied only at the critical point $\beta_c=0.5$ for $N-1$ modes. Furthermore, none of these modes reside in the midgap region, which is consistent with our theoretical prediction. Moreover, the bulk states we observe in the spectrum remain gapless only at $\beta_c$.

This conclusion applies in general, suggesting that we should not expect the presence of edge modes or localized modes in a chain consisting of identical nanoparticles. However, it is important to note that the existence of localized edge states is not a requirement for a topological phase transition. It is possible to have a topological phase transition without the presence of edge states. Therefore, while the absence of localized edge states in a chain of identical nanoparticles is expected, it does not undermine the possibility of a topological phase transition in such systems. The existence of a topological phase transition is determined by other topological characteristics and can be captured by appropriate topological invariants, which we will explore in subsequent section.  

\subsection{Topological Invariants: Zak Phase and Bulk Winding Number}
\label{Zakandwinding} 
\begin{figure}
\includegraphics[scale=1]{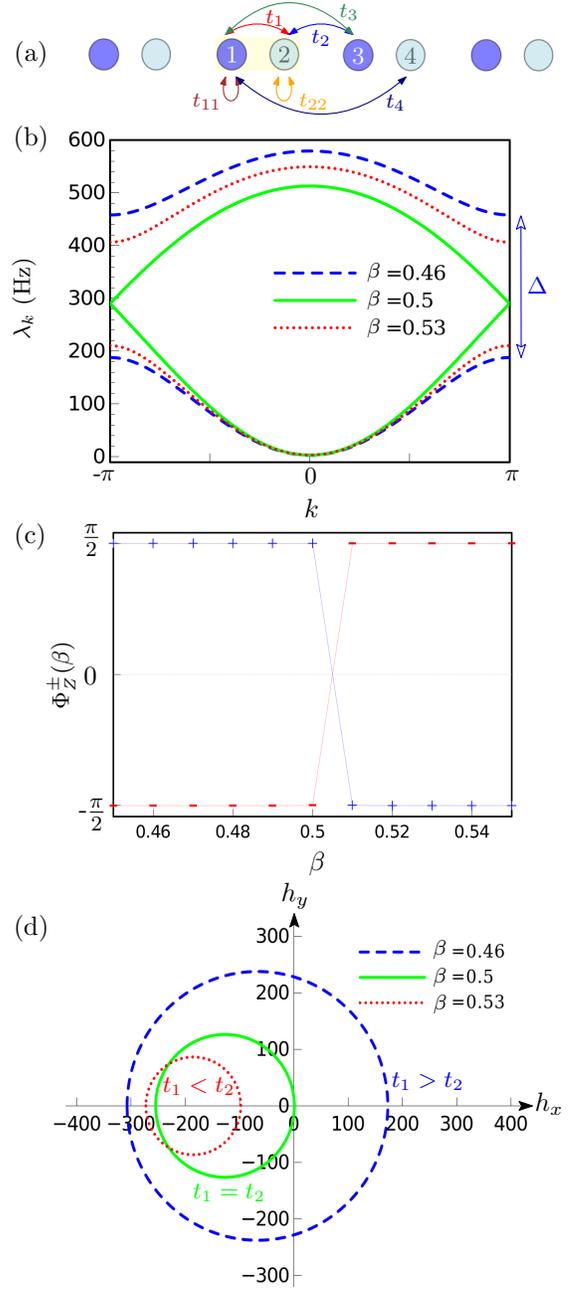}
\caption{(a) Schematic representation of a periodic chain with periodic boundary conditions. The inter-cell hopping is denoted as $t_{1}$, while the intra-cell hoppings are represented by $t_{2}$, $t_{3}$, and $t_{4}$. The on-site values are denoted as $t_{11}$ and $t_{22}$.
(b) Dispersion relations of the model described by Eq.~(\ref{eq20s}) for typical values of the control parameter $\beta$.
(c) Variation of the Zak phase with respect to the control parameter $\beta$.
(d) Trajectory of the endpoints of the vector $\vec h(k)$, which represents the bulk momentum-space response matrix described by Eq.~(\ref{eq19s}), traced on the $h_x$, $h_y$ plane as the wavenumber sweeps across the Brillouin zone, $k:-\pi\to \pi$.}
\label{FIGA4}
\end{figure}

To develop a comprehensive understanding of the topological phase transition, this section aims to investigate the topology of the system in reciprocal space. In order to mathematically represent the response matrix in reciprocal space, it is necessary to define the configuration of the unit. Referring to the configuration depicted in Fig.~(\ref{FIGA4}a), we can express the response matrix in reciprocal space in a concise manner as follows:
\begin{flalign}
\mathcal {\hat H} (k)=
\left[
\begin{array}{cc}
H_{11}+2H_{13}\cos(k) & H_{12}+H_{23}\text{e}^{ik}+H_{14}\text{e}^{-ik}\\ 
H_{12}+H_{23}\text{e}^{-ik}+H_{14}\text{e}^{ik} & H_{22}+2H_{24}\cos(k)
\end{array}
\right]
\label{eq18s}
\end{flalign}
represented compactly in notation as
\begin{flalign}
\mathcal {\hat H} (k)=h_x(k)\hat\sigma_x+h_y(k)\hat\sigma_y+h_z(k)\hat\sigma_z+h_0(k)\hat\sigma_0
\label{eq19s}
\end{flalign}
where $\hat\sigma_i~(i=x,y,z)$ is the Pauli matrix, $\hat\sigma_0$ is $2\times 2$ identity matrix, and $\vec h=h_x\hat i+h_y\hat j +h_z\hat k$ with
\begin{subequations}
\begin{eqnarray}
&&h_x(k)=t_{1}+t_{2}\cos(k)+t_{4}\cos(k),~~~~\\
&&h_y(k)=t_{2}\sin(k)-t_{4}\sin(k),~~~\\
&&h_z(k)=t_{11}/2-t_{22}/2,~~~~~\\
&&h_0(k)=t_{11}/2+t_{22}/2+2t_{3}\cos(k).~~~~
\end{eqnarray}
\end{subequations}
Based on the schematic representation of the hopping terms shown in Fig.~(\ref{FIGA4}a), the parameters $t_1 = H_{i,i+1}$, $t_2 = H_{i+1,i+2}$, $t_3 = H_{i,i+2} = H_{i+1,i+3}$, and $t_4 = H_{i,i+3} = H_{i+1,i+4}$ represent the hopping parameters that govern the power exchange between adjacent cells in the chain. These parameters control both the inter-cell and intra-cell hopping processes. Additionally, $t_{11} = H_{11}$ and $t_{22} = H_{22}$ correspond to the on-site terms of the response matrix and represent the cooling power contributions from the respective objects.

It is worth noting that the presence of the identity matrix $\sigma_0$ in Equation (\ref{eq19s}) results in a shift in the eigenvalues of the matrix $\mathcal{\hat H}(k)$. Additionally, in the given configuration, it is observed that $t_{11}$ and $t_{22}$ have similar magnitudes, resulting in the negligible contribution of the $h_z$ term. Conversely, we have $\min\{t_1,t_2\} \gg \max\{t_3,t_4\}$, which can be referred to as the nearest-neighbor interaction limit. In this limit, the model exhibits chiral symmetry, which implies that $\sigma_z \mathcal{\hat H} \sigma_z^\dagger = -\mathcal{\hat H}$.

The eigenvalues of the response matrix in Eq.~(\ref{eq19s}), which we refer to as thermal relaxation bands, can be expressed as:

\begin{equation}
\lambda^{\pm}_k=\pm|{\vec h}|+ h_0.
\label{eq20s}
\end{equation}

Here, $|{\vec h}|=\sqrt{h_x^2 + h_y^2 + h_z^2}$ is the magnitude of the vector ${\vec h}$. In Fig.~(\ref{FIGA4}b), the thermal relaxation bands for typical values of $\beta\in\{0.45,0.5,0.53\}$ are depicted. The thermal relaxation bands in the system exhibit a gap, denoted as $\Delta =\min \sqrt{h_x^2 + h_y^2 + h_z^2}$. This minimum gap condition plays a crucial role in determining the system's topological properties.  It can be observed that the two bands touch at $(k,\beta) = (\pm\pi,0.5)$, while a gap appears for $\beta \neq 0.5$. This information is significant as it allows us to define a topological invariant and investigate the transition of the system's topological phase.

The precise right eigenstate of the infinite system can be expressed as:

\begin{subequations}
\begin{eqnarray}
&&|\psi^+(k)\rangle=
\begin{pmatrix}
\cos(\frac{\theta}{2})~\text e^{-i\varphi}\\ 
\sin(\frac{\theta}{2})
\end{pmatrix},~~\\
&&|\psi^-(k)\rangle=
\begin{pmatrix}
\sin(\frac{\theta}{2})~\text e^{-i\varphi}\\ 
-\cos(\frac{\theta}{2})
\end{pmatrix}.~~
\end{eqnarray}
\label{eq21s}
\end{subequations}

In these equations, $\theta(k)=\arccos(h_z/|{\vec h}|)$ and $\varphi(k)=\arctan(h_y/h_x)$. Additionally, the left eigenvectors are denoted as $\langle\phi^\pm(k)|=|\psi^\pm(k)\rangle^\dagger$. The Zak phase provides a characterization of the topological properties of Bloch wave functions within the system. For the upper and lower bands, the geometric Zak phases are defined as follows:

\begin{equation}
\Phi_Z^\pm(\beta) = i\int_{-\pi}^{+\pi} \langle \phi^\pm(k) | \partial_k | \psi^\pm(k) \rangle dk.
\label{eq22s}
\end{equation}

In Equation (\ref{eq22s}), $\Phi_Z^\pm(\beta)$ represents the Zak phase for the respective upper ($+$) and lower ($-$) bands. The integral is taken over the range $-\pi$ to $+\pi$, and the terms $\langle \phi^\pm(k) |$ and $| \psi^\pm(k) \rangle$ denote the left and right eigenvectors, respectively, associated with the eigenstates of the system. 
To illustrate the computation of the Zak phase in our physical system with respect to the control parameter $\beta$, Fig.(\ref{FIGA4}c) is provided. It shows how the Zak phase changes as we vary $\beta$. This figure further demonstrates that the Zak phase transitions from $\pm\pi/2$ to $\mp\pi/2$ for the upper and lower bands, respectively, when $\beta$ crosses from values less than 0.5 to values greater than 0.5.

The eigenstates of the chain possess an internal structure that can be described by the direction of the vector $\vec{h}(k)$. In particular, due to the negligible value of $h_z$, as the momentum $k$ varies across the Brillouin zone from $-\pi$ to $\pi$, the endpoint of $\vec{h}(k)$ traces out a closed circular path on the $h_x$-$h_y$ plane. This circular path has a radius of $|t_1|$ and is centered at $(t_2+t_4, 0)$.

The topology of this circular path can be characterized by an integer known as the bulk winding number, denoted as $w$. The bulk winding number counts the number of times the circular path winds around the origin of the $h_x$-$h_y$ plane. In other words, it quantifies the number of revolutions the path completes around the origin as $k$ varies. In Fig.~(\ref{FIGA4}d), we can observe the behavior of the bulk winding number. For $\beta=0.53$, the winding number $w=0$, indicating that the circular path does not wind around the origin. On the other hand, for $\beta=0.46$, the winding number $w=1$, indicating that the circular path completes one full revolution around the origin. However, for $\beta=0.5$, the winding number is undefined, as the circular path coincides with the origin and does not wind around it.

\subsection{Adding defect to the system: asymmetric responce matrix}\label{defect} 

The absence of localized modes in this structure indicates that heat transfer in the system exhibits long-range characteristics, regardless of the value of $\beta$. Consequently, localizing or inhibiting the flow of heat within the system becomes challenging. To address this, our primary objective is to investigate methods that can induce the emergence of localized modes in the one-dimensional chain of nanoparticles. By achieving localized modes, we aim to enhance our ability to control heat transfer and manipulate thermal properties within the system.

To achieve this goal, we propose introducing a perturbation that breaks the symmetry of the system. This perturbation can take the form of a local effect or an asymmetry in the coupling between particles. In particular, we can explore the effect of varying the volumes of nanoparticles within the chain. By employing a simple approach, we consider a chain of nanoparticles with varying volumes. However, it is crucial to ensure the preservation of parity symmetry, as indicated by Eq.~(\ref{eqs16}). To simplify the analysis, we focus on a chain where the volumes of two specific particles depend on the parameter $\beta$. The volume configuration can be expressed as: 
\begin{equation}
  V_i(\beta)=\begin{cases}
    V_0=\text{cte}, & i\neq m,n.\\
    V(\beta), & i=m,n.
  \end{cases}
\end{equation} 
Here, $V_0$ represents the volume of the majority of nanoparticles in the chain, while $V(\beta)$ represents the volumes of the two specific nanoparticles located at positions $x_m$ and $x_n$. By introducing this strategy in the volume configuration, we break the translational symmetry of the system. Consequently, localized states can emerge around the nanoparticles with modified volumes. These localized states arise due to the confinement of the thermal flux within the region affected by the volume perturbation.

It is important to note that the specific form of the perturbation, $V(\beta)$, will determine the characteristics of the localized states, including their relaxation rate and spatial extent. The choice of the $V(\beta)$ should be made based on the desired properties of the localized states and the intended symmetry-breaking effect.

In this case, the response matrix exhibits asymmetric behavior. However, we can assess the existence and flatness of the edge band for $\beta<\beta_c$ by analyzing the derivative $\frac{d\lambda_e}{d\beta}$. By evaluating this quantity, we can gain insights into the behavior and characteristics of the edge mode in the system. From Eq.~(\ref{eqs17a}), we find $f_e(\beta)\simeq B V_e(\beta)$, where $B$ is a constant. Substituting this into Eq.~(\ref{eqs17b}) satisfies the flatness condition, as it reduces to $B\frac{dV_e}{d\beta}V_e-\frac{dV_e}{d\beta}BV_e=0$. Therefore, the existence of an edge state is possible if Eq.~(\ref{eqs17a}) holds true within the desired range of $\beta$. The next step is to determine a suitable function for $V_e(\beta)$.
\begin{figure}[t]
\includegraphics[]{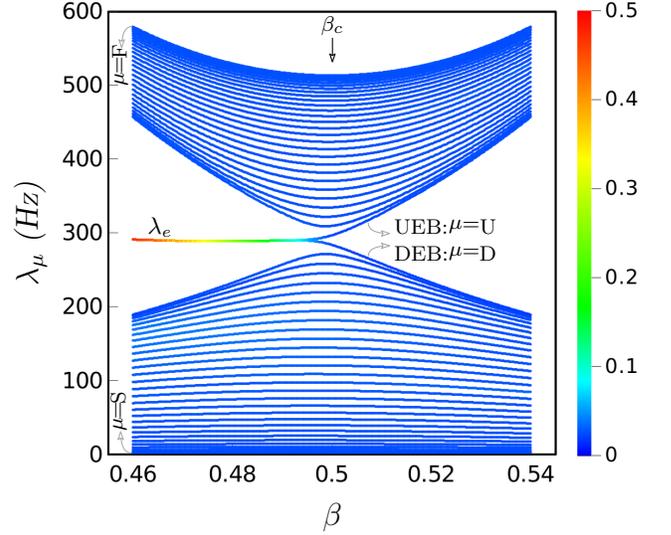}
\caption{Eigenvalue spectrum of a chain composed of SiC nanoparticles with a lattice constant of $d = 500$ nm and a total of $N = 60$ nanoparticles. The systerm characterized by a  asymmetric response matrix $\hat{H}$, where $V_{i\neq 1,60} = V_0 = \frac{4}{3}\pi 50^3 \text{ nm}^3$ and $V_1(\beta) = V_{60}(\beta) = \frac{4\pi}{3}[1250000\beta-561000]\text{ nm}^3$. The modes are sequentially labeled using Greek indices, and the significant bands include: $S$ (the slowest band with $\mu=1$), $F$ (the fastest band with $\mu=N$), $U$ (the higher edge band with $\mu=N/2+2$), and $D$ (the lower edge band with $\mu=N/2+1$). The colored inverse participation ratio (IPR) is depicted to illustrate the extended and localized eigenstates.}
\label{FIGA5}
\end{figure}
In general, we can express $f_e(\beta)$ as a polynomial for $\beta<\beta_c$. Since $V_e(\beta)$ is proportional to $f_e(\beta)$ within the desired range of $\beta$, we can assume, to first order in $\beta$, that $V_e(\beta)=V_m(\beta)=V_n(\beta)=a\beta+b$. Here, $a$ and $b$ are constants, and we can determine their values by minimizing the deviation of $\lambda_e$ in Eq.~(\ref{eqs17a}). By incorporating a properly defined $V_e(\beta)$, the conditions mentioned earlier hold true for $\beta<\beta_c$ and become invalid for $\beta>\beta_c$. Notably, introducing this asymmetry in nanoparticle volumes ensures the preservation of parity symmetry. Physically, the adiabatic change in volume denoted by $V(\beta)$, coupled with variations in the intra-cell separation distance, introduces a defect in the system. This defect retains most of the system's symmetries, except for translation symmetry. Consequently, two robust localized states emerge within the energy gap region. Thus, the topological phase transition in a chain of nanoparticles with an asymmetric response matrix is characterized by the emergence of localized modes.

Figure (\ref{FIGA5}) depicts the eigenvalue spectrum of the same nanoparticle chain discussed previously, but with the additional consideration that the volumes of the first and last particles are given by $V_1(\beta)=V_N(\beta)=\frac{4\pi}{3}[-1250000\beta+689000]$ nm$^3$. The coefficients $a=-\frac{4\pi}{3}1250000$ and $b=\frac{4\pi}{3}689000$ are chosen to ensure that the constraints in Eq.~(\ref{eqs17}) hold true for values of $\beta$ less than 0.5. It is apparent from the spectrum that a topological edge band is present, and a phase transition occurs at the critical point $\beta_c=0.5$. 
The specific behavior and properties of these localized modes depend on the details of the defect itself, such as its position, volume $V_0$, lattice constant $d$, and the specific form of the function $V(\beta)$ introduced by the variation in $\beta$ and intra-cell separation distance. The function $V(\beta)$ controls the adiabatic change in the system's parameters, affecting the temperature field behavior in the vicinity of the defect. The localization of this modes (known as defect modes) is a result of the broken translational symmetry caused by the defect, leading to a modification of the system's periodicity. This disruption causes the eigenstates associated with the defect to become localized within the band gap, resulting in the emergence of the defect modes.
\begin{figure}
\includegraphics[scale=0.9]{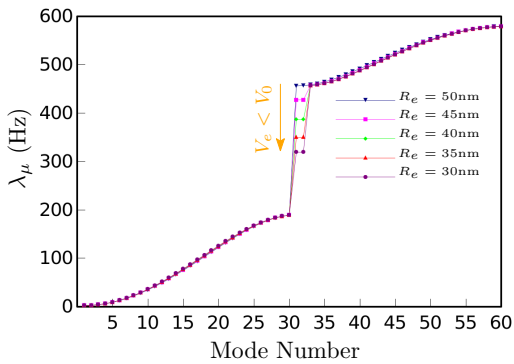}
\caption{The eigenvalue spectrum in a chain consisting of 60 nanoparticles at $\beta = 0.46$. The spectrum is presented for different values of $R_e$, representing the radius of the nanoparticle at position $(m,n) = (1,60)$, while all other particles maintain a constant radius of $R_0 = 50$~nm. Initially, the system displays a symmetric response matrix when $R_e = R_0$. However, as the value of $R_e$ decreases, the system undergoes a transition to an asymmetric response matrix.}
\label{FIGA6}
\end{figure}
As shown in Fig.~(\ref{FIGA5}), the edge band with a higher eigenvalue in the topologically trivial phase ($\beta>0.5$) is labeled as $U$, while the other band with a lower eigenvalue is labeled as $D$. In the subsequent discussions, their corresponding eigenmodes are referred to as the "upper" edge mode $\psi_U(x)$ and the "lower" edge mode $\psi_D(x)$, respectively. Moreover, the eigenmode in the spectrum with the lowest (highest) eigenvalue is labeled as $S$ ($F$). According to Ref.~\cite{PhysRevLett126193601}, the first mode $\boldsymbol{\psi}_S$ predominates in the thermalization process when excited in a collection of particles. In the nontrivial topological phase of the system ($\beta<0.5$), the edge modes become localized at both ends of the chain. However, it will be demonstrated in the subsequent section that the topologically localized modes can be positioned at any arbitrary point within the chain, denoted as $(x_m,x_n)$, with the condition $m=N+1-n$.

In order to investigate the underlying mechanism behind the emergence of localized edge states as we transition from a symmetric to an asymmetric response matrix, we analyze the eigenvalue spectrum of a chain of length $N=60$ with $R_i = R_0 = 50~\text{nm}$ for $\beta = 0.46$, as shown in Fig.~(\ref{FIGA6}). By varying the radii of particles $1$ and $60$, denoted as $R_e$, the response matrix undergoes a transition from symmetric to asymmetric. When $R_e = R_0$, corresponding to a symmetric response matrix, the system exhibits a spectrum with a finite  gap and no localized states as we observed in Fig.~(\ref{Figure.1}b). However, as $R_e$ is decreased below $R_0$, the response matrix becomes asymmetric.

In this asymmetric regime, two defect modes emerge from the upper bulk band and move into the gap region. Notably, these modes reach the midgap for nonzero values of $R_e$. Remarkably, the introduction of an asymmetric response matrix enables the existence of localized modes without modifying the coupling configuration within the bulk of the system. In our subsequent analysis, we will demonstrate that these modes are exponentially confined around the defect position, highlighting their localized nature.

\subsection{Robustness of the edge states}
\label{robust} 
The robustness of topological edge states is a fundamental characteristic that ensures their protection against perturbation and imperfections in the system. These edge states emerge in specific topological phases of matter, such as topological insulators or topological superconductors, where excitations are confined to the boundaries or interfaces of the material. As illustrated in Fig.~(\ref{FIGA5}), the presence of a gap in the eigenvalue spectrum of the system with an asymmetric response matrix distinguishes the edge states from the bulk states. This gap serves as a protective barrier, preventing scattering or hybridization between the edge states and the bulk states. Consequently, the edge states exhibit resilience against local perturbations that do not result in the closure of the gap.

To demonstrate the robustness of topological edge states in the proposed system, we investigate the effects of perturbation on the eigenvalue spectrum of the response matrix. The perturbation is introduced by adding uncorrelated random numbers with a Gaussian distribution and a standard deviation of $\sigma$ to the elements of the response matrix in real space. The parameter $\sigma$ quantifies the strength of local perturbation, encompassing various factors such as displacement, removal, or changes in material or particle volume. In this section, the presented results are obtained for $\beta=0.46$, and statistical significance and reliability are ensured by averaging over 500 realizations.
\begin{figure}
\includegraphics[scale=1]{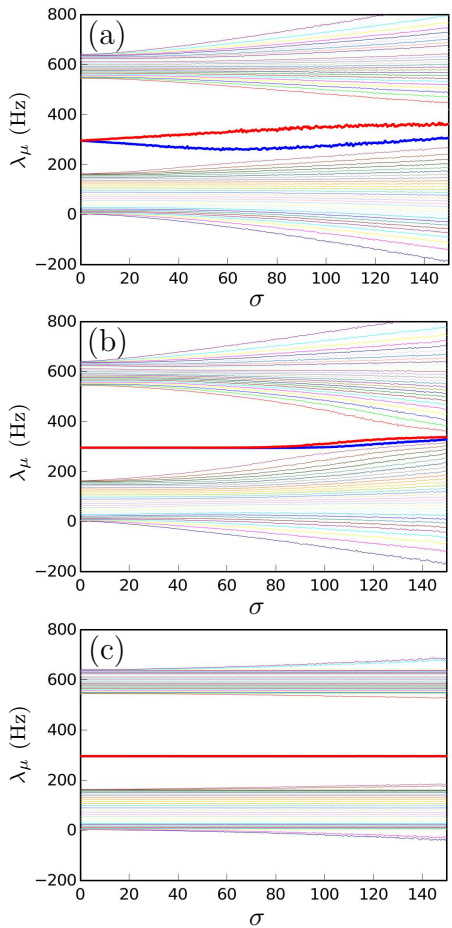}
\caption{Effect of (a) diagonal, (b) off-diagonal, and (c) Mirror-reflection perturbation  on the eigenvalue spectrum in a finite chain with $N = 61$ particles and $(m,n)=(1,60)$, as a function of the perturbation strength $\sigma$.}
\label{FIGA7}
\end{figure}
As shown in Fig.(\ref{FIGA7}a), the introduction of diagonal perturbation breaks the chiral symmetry, leading to the loss of topological protection for the edge modes within the system. This disruption results in the mixing of bulk and edge states, as evidenced by the overlap of their eigenvalues. Figure(\ref{FIGA7}b) illustrates the impact of off-diagonal perturbation on the edge states. In this case, the localized edge states exhibit resilience to the specific perturbation, maintaining their separation from the bulk states. However, the perturbation parameter $\sigma$ noticeably affects the eigenvalues of the bulk bands. We observe in both cases that as the magnitude of perturbation approaches the band gap, this effect becomes more pronounced and can ultimately close the gap, causing the edge states to merge with the bulk states.

The question at hand is whether there exists a perturbation that can be introduced into the model without closing the energy gap between the bulk bands. The model exhibits a distinct topological phase characterized by the presence of protected edge states when the relaxation gap is present. To preserve this gap while introducing perturbations, it is crucial to maintain the system's underlying topological properties and preserve the relevant symmetries.
In this context, our focus lies on perturbations that preserve the Mirror-reflection symmetry of the system, as illustrated in Figure (\ref{FIGA7}c). This specific perturbation is carefully designed to uphold the mirror reflection symmetry on the perturbed hoppings, while simultaneously avoiding the closure of the band gap. Consequently, the introduced perturbation safeguards the topological invariant of the system, which, in this case, is represented by the winding number. By maintaining the Mirror-reflection symmetry, the perturbation effectively preserves the system's topological phase and ensures the persistence of the associated protected edge states.

\begin{figure*}[]
\includegraphics[]{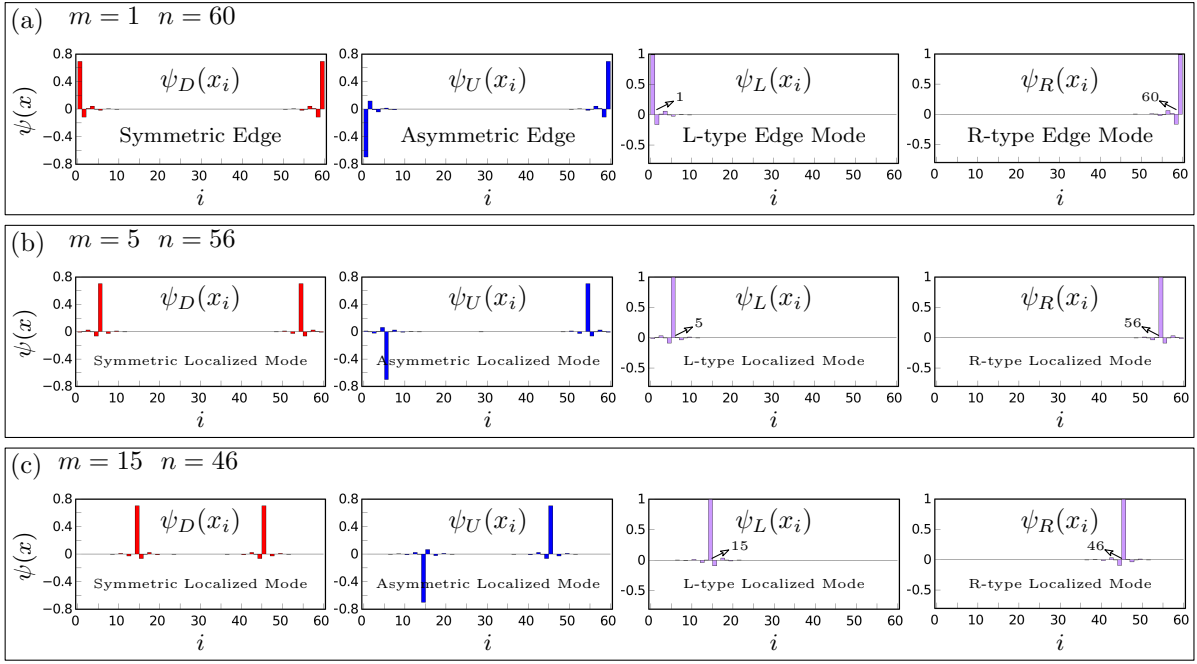}
\caption{The localized states and corresponding L-type and R-type states for the asymmetric chain with $d=500$~nm and $N = 60$, in its nontrivial phase for $\beta=0.46$. The volume of particles are  $V_{i\neq m,n}=\frac{4}{3}\pi 50^3$~nm$^3$ and $V_m(\beta)=V_{n}(\beta)=\frac{4\pi}{3}[1250000\beta-561000]\text{nm}^3$. (a) Localized modes on particles 1 and 60, (b) Localized modes on particles 5 and 56, (c) Localized modes on particles 15 and 46.}
\label{FIGA8}
\end{figure*} 
\section{Localized modes of the System with asymmetric Responce Matrix}

\subsection{ L-type and R-type Eigenmodes }\label{LR}

As discussed earlier, $\psi_U(x)$ and $\psi_D(x)$ are simultaneous eigenstates of $\hat{H}$ and $\hat{\Pi}$. They are degenerate states in the topologically nontrivial phase of the system, but not in the trivial phase. Hybridized modes can be formed as follows:
\begin{subequations}
\label{eqs18}
\begin{eqnarray}
\label{eqs18a}
&&\psi_L(x)=\frac{1}{\sqrt 2}[\psi_U(x)-\psi_D(x)],\\
\label{eqs18b}
&&\psi_R(x)=\frac{1}{\sqrt 2}[\psi_U(x)+\psi_D(x)],
\end{eqnarray}
\end{subequations}
\begin{figure}[]
\includegraphics[scale=0.9]{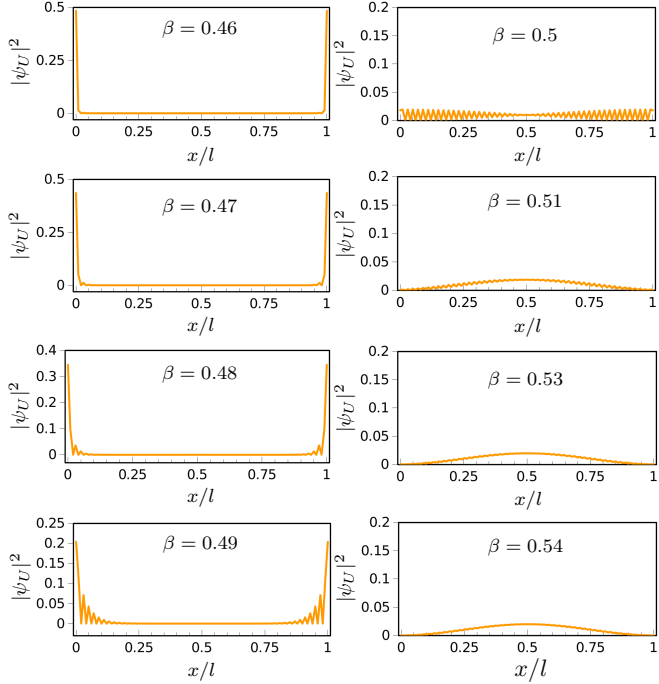}
\caption{ Probability distribution of $\psi_U(x)$ for $\beta\in[0.46,0.54]$ in an asymmetric chain of $N=60$ NPs, where $V_{i\neq 1,60}=\frac{4}{3}\pi 50^3$~nm$^3$ and $V_1(\beta)=V_{60}(\beta)=\frac{4\pi}{3}[1250000\beta-561000]\text{nm}^3$.}
\label{FIGA9}
\end{figure}
These modes have weight distributions $P_L=P_R=0.5[P_U+P_D]$ and are referred to as L-type and R-type states, respectively. It is important to note that these states are not parity eigenstates; however, they are eigenstates of $\hat{H}$ in the topologically nontrivial phase:
\begin{subequations}
\label{eqs19}
\begin{eqnarray}
\label{eqs19a}
&&\hat H \psi_L(x)=\lambda_e\psi_L(x),\\
\label{eqs19b}
&&\hat H \psi_R(x)=\lambda_e\psi_R(x).
\end{eqnarray}
\end{subequations}
Furthermore, the states described by Equations (\ref{eqs18a}) and (\ref{eqs18b}) are mostly concentrated on the left-hand and right-hand localization sites, respectively. However, in the trivial phase, they are not eigenstates of $\hat{H}$ and are extended over the entire chain.

The localized modes and their corresponding L-type and R-type states are depicted in Fig.~(\ref{FIGA8}) for $\beta=0.46$. The positions of the localized states are chosen to maintain x-directional mirror symmetry with respect to $l/2$ by setting $V_{i\neq m,n}=V_0$ and $V_m=V_{n}=V(\beta)$. Each panel in Figure (\ref{FIGA8}) represents a typical value of $(m,n)$, with $n=N+1-m$. It is evident that the corresponding L-type and R-type states in each case are located on the $m$th and $n$th sites, respectively. We note that the L-type and R-type states can be considered as pure states in the limit of $\beta\ll \beta_c$.

 \begin{figure}[]
\includegraphics[scale=0.92]{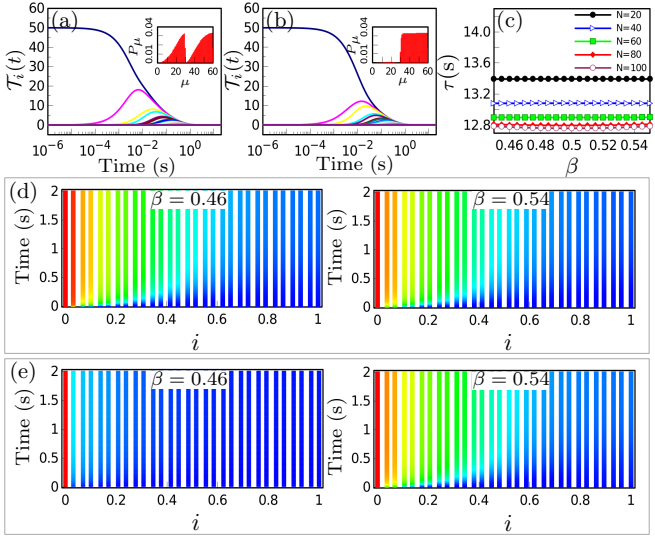}
\caption{ Evolution of temperatures in a chain with symmetric response matrix and $N=60$ nanoparticles with volumes $V_{i}=\frac{4}{3}\pi 50^3$~nm$^3$ for (a) $\beta=0.46$ and (b) $\beta=0.54$. As an initial temperature condition, only the $1$-th particle is heated up to $350$~K, while all other particles being initially at room temperature $300$~K. (c) Thermalization time for the initial temperature state $\Delta T_i(0)=50\delta_{i,1}$, in a chain of NPs with symmetric response matrix and length $N\in\{20,40,60,80,100\}$. The spatio-temporal evolution of the temperature field, denoted as $\Delta T(x_i,t)$, is studied in a chain consisting of $N=30$ particles. The first particle is maintained at a constant temperature $T_1 = 350$ K, while all other particles are initially at room temperature $300$ K: (d) The temperature field $\Delta T(x_i,t)$ in a chain with a symmetric response matrix, corresponding to the setup shown in Fig.~(\ref{Figure.1}b).  (e) The temperature field $\Delta T(x_i,t)$ in a chain with an asymmetric response matrix, following the configuration depicted in Fig.~(\ref{Figure.1}c).}
\label{FIGA10}
\end{figure}

\subsection{Localized Vs Extended Modes}  \label{sec44}
To further investigate the physical characteristics of the model illustrated in Fig.~(\ref{FIGA5}), we utilize the inverse participation ratio (IPR). The color bars on the right side of the figure indicate the IPR values associated with the modes in the system featuring an asymmetric response matrix. Notably, we observe that the edge modes within the topologically nontrivial phase consistently display the highest IPR values. Remarkably, the IPR of these modes remains the largest irrespective of the specific topological phase of the system.

Figure (\ref{FIGA9}) illustrates the variation in the profile of $|\psi_U(x)|^2$ across the critical point $\beta_c=0.5$ for the same configuration as depicted in Figure (\ref{FIGA5}), considering both the topologically trivial and nontrivial phases. At $\beta=0.46$, the profile is entirely localized at $x=0$ and $x=l$. However, as $\beta$ increases, the contribution of the edge state at $x=0$ and $x=l$ diminishes. This delocalization is supported by the decrease in the IPR of $\psi_U(x)$ in Figure (\ref{FIGA5}). Upon crossing the critical point, the state becomes extended, indicating the occurrence of an edge-bulk transition. In the topologically trivial phase, we observe a highly uniform profile within the chain, and $\psi_U(x)$ becomes delocalized over the entire system.


\section{Temporal Evolution of Temperatures in Chain with symmetric Response Matrix}\label{temporalinhamiltonian}

In Fig.~(\ref{FIGA10}), we investigate the temperature evolution in a chain comprising $N=60$ nanoparticles with volumes $V_{i}=\frac{4}{3}\pi (50^3)$ nm$^3$. The chain exhibits a symmetric response matrix. We examine two cases: (a) $\beta=0.46$ and (b) $\beta=0.54$. Initially, only the first particle (particle $1$) is heated to a temperature of $350$ K, while all other particles are at room temperature ($300$ K). In both cases, we observe that particle number 1 undergoes heat exchange with other particles, eventually reaching thermal equilibrium. This equilibrium state is attained after multiple thermalization steps within the system.

Considering the information presented in Fig.(\ref{FIGA4}c), which illustrates a topological phase transition in the system as the parameter $\beta$ changes, while no localized states are observed, and taking into account the inset, which indicates the involvement of a significant number of modes in the temperature evolution, we can reasonably expect a prolonged thermalization time for both scenarios. These results align with the long-range nature of heat exchange in the SSH chain of identical nanoparticles, as reported in Ref.\cite{PhysRevB102115417}.

The dependence of thermalization time on the chain length is shown in Fig.~(\ref{FIGA10}c) as a function of $\beta$. The plot reveals a decreasing trend in thermalization time as the chain length increases, eventually reaching a saturation point. On the other hand, the variation of parameter $\beta$ does not exert a significant influence on the thermalization time of the system. However, it is worth noting that in longer chains, the thermalization time reaches its minimum value at $\beta = 0.5$. Once again, we can infer that the absence of localized states accounts for the lack of dependence of the thermalization time on the parameter $\beta$ in a chain of identical nanoparticles.

Finally, we compare the spatio-temporal evolution of the temperature field in a chain between a system with a symmetric and asymmetric response matrix. In both scenarios, the first particle is maintained at a fixed temperature of $350$ K, while the temperatures of the rest of the chain are initially set to the room temperature of $300$ K. The results are presented for $\beta = 0.46$ and $\beta = 0.54$. Figures (\ref{FIGA10}d) and (\ref{FIGA10}e) depict the results for the system with a symmetric and asymmetric response matrix, respectively.

In the asymmetric case, we consider a chain length of $N = 30$ and set the values of $(m, n) = (1, 30)$ to ensure the presence of topologically non-trivial edge state. Interestingly, we observe that the thermal energy becomes localized only in the topologically non-trivial phase of the system with an asymmetric response matrix. This localization arises due to the interplay between the asymmetric effects and the system's topological properties, leading to the confinement of thermal energy in specific regions of the chain.

In contrast, in the spatio-temporal evolution of the temperature field in the chain with a symmetric response matrix, no such localization is observed. The absence of localization in the symmetric case is consistent with the behavior of a system without topological protection, where thermal energy can freely propagate throughout the chain without being confined to specific regions.

These findings highlight the crucial role of the asymmetric response matrix in inducing the emergence of localized states and the distinct behavior of the system compared to the symmetric case. Moreover, they provide further evidence for the connection between asymmetric physics, topological properties, and the spatial distribution of thermal energy in the studied chain system.

\section*{reference}
\bibliographystyle{unsrt}

\end{document}